\documentclass[aps,prd,twocolumn]{revtex4-2}
\usepackage[
colorlinks=true,
filecolor=black,
anchorcolor=blue,
linkcolor=blue,
citecolor=cyan,  
urlcolor=cyan,
linktocpage=true,
plainpages=false,
breaklinks=true,
pdfstartview=FitH
]{hyperref}

\usepackage{amsmath}
\usepackage{graphicx}    
\usepackage{xcolor}
\usepackage[normalem]{ulem}

\begin{document}
\title{Alternative LISA-TAIJI networks: Detectability of the Parity Violation \\ in Stochastic Gravitational Wave Background } 

\author{Ju Chen$^{1,2}$}
\email{chenju@ucas.ac.cn}
\author{Chang Liu$^{3}$}
\email{Corresponding author: liuchang@yzu.edu.cn}
\author{Yun-Long Zhang$^{4,5}$}
\email{Corresponding author: zhangyunlong@nao.cas.cn}
\author{Gang Wang$^{6,7,8}$}
\email{Corresponding author: gwanggw@gmail.com}

\affiliation{$^{1}$International Center for Theoretical Physics Asia-Pacific (ICTP-AP), University of Chinese Academy of Sciences, Beijing 100190, China}
\affiliation{$^{2}$Taiji Laboratory for Gravitational Wave Universe (Beijing/Hangzhou), University of Chinese Academy of Sciences, Beijing 100049, China}
\affiliation{$^{3}$Center for Gravitation and Cosmology, College of Physical Science and Technology, Yangzhou University, Yangzhou, 225009, China}
\affiliation{$^{4}$National Astronomical Observatories, Chinese Academy of Sciences, Beijing, 100101, China}
\affiliation{$^{5}$School of Fundamental Physics and Mathematical Sciences,  Hangzhou   Institute for Advanced Study, UCAS, Hangzhou 310024, China}
\affiliation{$^{6}$Institute of Fundamental Physics and Quantum Technology, Ningbo University, Ningbo, 315211, China}
\affiliation{$^{7}$Department of Physics, School of Physical Science and Technology, Ningbo University, Ningbo, 315211, China}
\affiliation{$^{8}$Shanghai Astronomical Observatory, Chinese Academy of Sciences, Shanghai, 200030, China}

\date{\today}
\allowdisplaybreaks

\begin{abstract}

The detection of parity violation in the isotropic stochastic gravitational wave background (SGWB) will serve a crucial probe for new physics, particularly in parity-violating theories of gravity. The joint observations by the planned space-borne gravitational wave detectors, LISA and TAIJI, will offer a unique opportunity to observe such effects in the millihertz (mHz) band. This study evaluates the detectability of parity violation in the SGWB using two network configurations: LISA-TAIJIp and LISA-TAIJIm. The former configuration consists of LISA (inclined at $+60^\circ$ relative to the ecliptic plane) and TAIJIp (also inclined at $+60^\circ$), while the latter network pairs LISA with TAIJIm (inclined at $-60^\circ$).
Our analysis demonstrates that the sensitivity of the LISA-TAIJIm network to parity violation in the SGWB is approximately one order of magnitude greater than that of the LISA-TAIJIp network at lower frequencies. To quantify the performance of the two networks, we evaluate the signal-to-noise ratios for different spectral shapes, including power-law, single-peak, and broken power-law models, and estimate parameter determination using the Fisher information matrix. The results confirm that LISA-TAIJIm outperforms LISA-TAIJIp in detecting the SGWB with circular polarization components, offering a superior opportunity to test parity-violating gravitational constraints on various mechanisms in the mHz band.
\end{abstract}


\maketitle

\section{Introduction}
The stochastic gravitational wave background (SGWB) is expected to arise from the superposition of gravitational waves (GWs) produced by numerous independent sources. Various astrophysical processes, such as close compact binaries and rotating neutron stars, can generate the SGWB. Additionally, cosmological processes, including inflation and first-order phase transitions in the early universe, can also contribute to the SGWB. A recent review of the different sources of the SGWB is provided in \citep{Christensen2019}. Detecting the SGWB would offer valuable insights into the evolution of astrophysical sources as well as the history of the early Universe.

The SGWB is typically assumed to be unpolarized, attributed to the stochastic and uncorrelated nature of its generation process. However, the presence of parity violation in gravity would modify the generation and propagation of GWs, leading to a circularly polarized background. The (non-)detection of such a background would provide a new approach to understanding gravity and help place constraints on various parity-violating mechanisms \citep{Callister:2023tws}. Many parity-violating effects have been explored concerning the generation of the SGWB in the early Universe, such as leptogenesis \citep{Alexander2006, Maleknejad:2016qjz}, Chern-Simons coupling during inflation \citep{Satoh2008,Adshead2012},
axion-like mechanism  
during the radiation-dominated era \citep{Machado:2018nqk,Machado:2019xuc,Maleknejad:2024vvf}.
Additionally, helical turbulence in the primordial plasma during first-order phase transition, induced by primordial magnetic fields, can also generate a circularly polarized SGWB \citep{Kahniashvili2005,RoperPol2020}. 
Apart from the generation process, violations of parity could also lead to birefringence in the propagation of GWs \citep{Callister:2023tws,Califano:2023aji,Li:2024fxy}, causing the left-hand and right-hand circular polarization modes to evolve differently, thereby resulting in a net circular polarization upon measurement.

The search for circular polarization in the SGWB has been carried out across various frequency bands using different GW detectors. For ground-based detectors, the sensitivity is currently not sufficient to obtain a significant constraint on the circular polarization \citep{Martinovic2021,Jiang2023a}, but measurements are expected in the next few years \citep{Omiya2023a}. For space-borne detectors, the detection limits can potentially reach $10^{-12}$ in terms of fractional energy density with ten years of observation \citep{Seto2020,Orlando2021}.
Furthermore, studies have shown that Pulsar Timing Arrays (PTAs) are insensitive to the circular polarization of an isotropic background. Nevertheless, measurements of the circular-polarization anisotropy are theoretically possible \citep{Kato2016,Belgacem2020a}.

LISA (Laser Interferometer Space Antenna) is a space-borne gravitational wave detector scheduled for launch in the late 2030s, with a planned nominal mission duration of 4 years, extendable to 10 years \citep{LISA:2017pwj,Colpi:2024xhw}. TAIJI, similar to LISA, is expected to launch during the same period \citep{Ruan2018}.
Both LISA and TAIJI are heliocentric detectors consisting of a triangular constellation formed by three spacecraft, with arm lengths of 2.5 million kilometers and 3.0 million kilometers, respectively. LISA is planned to trail the Earth at an angle of $\beta \sim 20^\circ$, while TAIJI leads the Earth at the same angle.
To maintain stable constellation formation, the constellation planes of both detectors are designed to be inclined by approximately $\pm 60^\circ$ relative to the ecliptic plane \citep{Dhurandhar2005}.

Considering the overlap of mission schedules, a network between LISA and TAIJI is expected to form \citep{Ruan2020a}, offering significant advantages. Previous studies have demonstrated that such a network can greatly improve the sky localization of massive binary black holes \citep{Ruan2020a, Wang2021c} and enhance the observation of stellar-mass binary black holes \citep{Chen2021c,Zhao2023a}. A recent overview can be found in \citep{Cai2023}.
Notably, the circular polarization of an isotropic gravitational wave (GW) background, which is the primary focus of this work, cannot be detected with a single planar detector \citep{Seto2006,Seto2007}. In previous work \citep{Chen:2024ikn}, we derived the general formula for the correlation analysis of space-based detectors in the long-wavelength limit and found that sensitivity can be significantly enhanced by optimizing the network configuration. In this paper, we focus on the alternative LISA-TAIJI networks proposed in \citep{Wang2021c}, evaluate the detectability of a parity-violating SGWB with realistic orbits and a more robust TDI scheme, and compare the parameter uncertainties with common power density spectrum models.

In the following, we introduce the alternative LISA-TAIJI networks in Section~\ref{sec:network} and the parity of the SGWB in Section~\ref{sec:SGWB}. We analyze the effective overlap reduction functions and the power-law integrated sensitivity curve in Section~\ref{sec:result}, and determine the parameters of the SGWB with different energy density spectra in Section~\ref{sec:parameters}. Finally, we present the conclusion and discussion in Section~\ref{sec:conclusion}.

\section{LISA-TAIJI Networks}
\label{sec:network}

\begin{figure}[htp]
\includegraphics[width=0.425\textwidth]{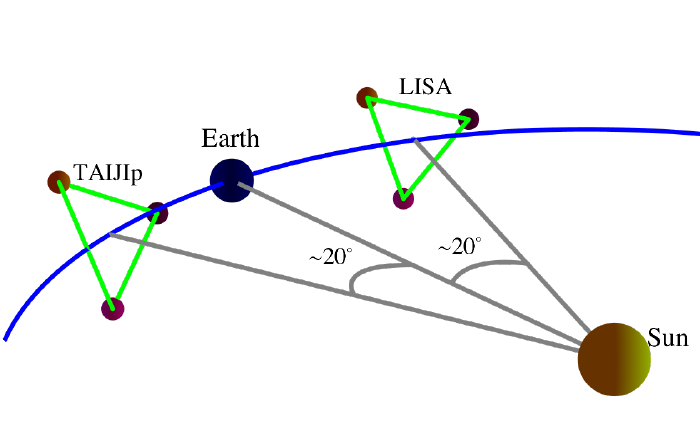}
\includegraphics[width=0.425\textwidth]{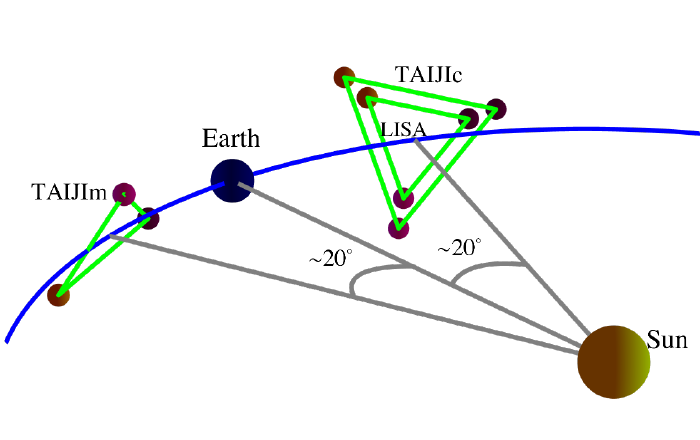}
 \caption{Configurations of alternative LISA-TAIJI networks. The LISA is designed to trail the Earth by $\sim 20^\circ$, with its triangular constellation inclined at $+60^\circ$ relative to the ecliptic plane. For the TAIJI mission, three alternative orbital formations have been proposed \citep{Wang2021c}: TAIJIp, where the constellation leads the Earth by $\sim 20^\circ$ with a $+60^\circ$ inclination; TAIJIm, where the constellation leads the Earth by $\sim 20^\circ$ with a $-60^\circ$ inclination; and TAIJIc, where the constellation is coplanar with LISA.}
 \label{fig:orbit}
\end{figure}

To evaluate the optimal orbital configuration of TAIJI for the network, \citet{Wang2021c} introduced three alternative TAIJI orbits: TAIJIp, TAIJIm, and TAIJIc, as depicted in Fig.~\ref{fig:orbit}. The orbits of TAIJIp and TAIJIm both lead the Earth by $\sim 20^\circ$, with their constellation planes inclined at $+60^\circ$ and $-60^\circ$ relative to the ecliptic plane, respectively. The TAIJIc is colocated and coplanar with the LISA.
It has shown that both LISA-TAIJIp and LISA-TAIJIm are suitable for observing massive black hole binary coalescence, as the large separation significantly improves the source sky localization. Additionally, LISA-TAIJIm can also achieve higher parameter resolutions for the MBBHs due to its more misaligned configuration \cite{Wang2021c}. For an isotropic background, the sensitivities of LISA-TAIJIp and LISA-TAIJIm vary with frequencies but are generally comparable for the typical spectral shapes \citep{Wang2021d}. In this work, we exclude the coplanar case LISA-TAIJIc, which is insensitive to the isotropic circularly polarized SGWB.

For space-borne interferometers like LISA and TAIJI, time-delay interferometry (TDI) is essential for the suppression of laser frequency noise \citep[and references therein]{1999ApJ...527..814A,2000PhRvD..62d2002E,2003PhRvD..67l2003T,Vallisneri:2005ji,Tinto:2020fcc}. Various TDI channels can be constructed by combining measurements from different arm links. In this work, instead of using the Michelson TDI configuration, we employ the more robust hybrid Relay TDI scheme to perform evaluations \citep{Wang:2011tlj,Wang:2024alm,Wang:2024hgv,Wang:2020pkk,Muratore:2020mdf}. The corresponding orthogonal channels A and E are utilized for the specific calculations \citep{Prince2002}. Numerical calculations are carried out using SATDI \citep{Wang:2024ssp}. The full expressions of the noise power spectral density (PSD) for the hybrid Relay TDI configuration can be found in \citep{Wang:2024alm}. Under the equal arm-length assumption, the noise PSD of channels A and E can be simplified as
\begin{equation}
\begin{aligned}
N_A = N_E &= N_{\rm acc} \big( 56 -12 \cos x - 36 \cos 3x  - 12 \cos 4x \\ &
   - 20 \cos 5x + 16 \cos 6x + 4 \cos 7x + 4 \cos 8x \big) \\ &
+ N_{\rm oms} \big( 20 - 4 \cos x - 6 \cos 2x - 8 \cos 3x\\&
- 4 \cos 4x - 6 \cos 5x + 6 \cos 6x + 2 \cos 7x \big),
\end{aligned}
\end{equation}
where $x = 2\pi f L/c$, $N_{\rm oms}$ is the noise budget of the optical measurement system, and $N_{\rm acc}$ is the acceleration noise of the test mass for the corresponding detectors. The noise budgets of the LISA and TAIJI detectors are assumed to be  
\begin{eqnarray}
& N_{\rm acc} &= A^2_{\rm acc} \left[1 + \left(\frac{0.4 \rm mHz}{f}\right)^2\right]\left[1 + \left(\frac{f}{8 \rm mHz}\right)^2\right], \\
& N_{\rm oms} &= A^2_{\rm oms} \left[1 + \left(\frac{2 \rm mHz}{f}\right)^4\right].
\end{eqnarray}
For LISA, the noise amplitudes are set to $A_{\rm oms} = 10 \, \mathrm{pm}/\sqrt{\mathrm{Hz}}$ and $A_{\rm acc} = 3 \, \mathrm{fm/s^2}/\sqrt{\mathrm{Hz}}$ \citep{LISA:2017pwj}.  
For TAIJI, the noise budgets are $A_{\rm oms} = 8 \, \mathrm{pm}/\sqrt{\mathrm{Hz}}$ and $A_{\rm acc} = 3 \, \mathrm{fm/s^2}/\sqrt{\mathrm{Hz}}$ \citep{Luo:2019zal}.

\section{Parity of gravitational wave Backgrounds}
\label{sec:SGWB}

The stochastic background can be viewed as a superposition of plane waves from all possible directions and frequencies \citep{Maggiore2007}:
\begin{equation}
h_{ab}(t, \vec{r}) = \sum_{P=+, \times} \int_{-\infty}^{\infty} df \int d^2\hat{k} ~ \tilde{h}_P(f, \hat{k}) e_{ab}^P(\hat{k}) e^{i 2\pi f \left( t - \hat{k} \cdot \vec{r} / c \right)},
\end{equation}
where $e^P_{ab}(\hat{k})$ is the polarization tensor for a given propagation direction $\hat{k}$, and $\tilde{h}_P(f, \hat{k})$ are the Fourier coefficients corresponding to polarization $P$ and direction $\hat{k}$.

Here, we only consider the two polarizations in general relativity, $P = +, \times$, and following \citep{Allen1999}, the polarization tensors are defined as
\begin{equation}
e^+_{ab}(\hat{k}) = \hat{m}_a \hat{m}_b - \hat{n}_a \hat{n}_b, \quad
e^\times_{ab}(\hat{k}) = \hat{m}_a \hat{n}_b + \hat{n}_a \hat{m}_b,
\end{equation}
where $\hat{m}$ and $\hat{n}$ are orthogonal unit vectors in the plane perpendicular to the propagation direction $\hat{k}$. The vectors $\hat{k}$, $\hat{m}$, and $\hat{n}$ are given by
\begin{equation}
\begin{aligned}
  \hat{k} &= \cos\phi \sin\theta ~ \hat{x} + \sin\phi \sin\theta ~ \hat{y} + \cos\theta ~ \hat{z},\\
  \hat{m} &= \sin\phi ~ \hat{x} - \cos\phi ~ \hat{y},\\
  \hat{n} &= \cos\phi \cos\theta ~ \hat{x} + \sin\phi \cos\theta ~ \hat{y} - \sin\theta ~ \hat{z},
\end{aligned}
\end{equation}
where $(\theta, \phi)$ are the standard polar and azimuthal angles in spherical coordinates. With these definitions, the polarization tensors satisfy the orthogonality relation $e^P_{ab}(\hat{k}) e^{P'}_{ab}(\hat{k})=2\hat{\delta}^{PP'}$.
Here, $\hat{\delta}^{PP'}$ is the Kronecker delta.

In the following, we assume the background is stationary and isotropic, with a possible nonzero net polarization. In this case \citep{Seto2006}:
\begin{widetext}
\begin{equation}
\label{eq:stocks}
\begin{pmatrix}
  \langle \tilde{h}_+(f, \hat{k}) \tilde{h}_+^*(f', \hat{k}')\rangle & \langle \tilde{h}_+(f, \hat{k}) \tilde{h}_\times^*(f', \hat{k}')\rangle \\
   \langle \tilde{h}_+^*(f, \hat{k}) \tilde{h}_\times(f', \hat{k}')\rangle & \langle \tilde{h}_\times(f, \hat{k}) \tilde{h}_\times^*(f', \hat{k}')\rangle
\end{pmatrix} 
= \frac{1}{2}\delta(f-f') \frac{\delta^2(\hat{k}-\hat{k}')}{4\pi}
\begin{pmatrix}
I(f) + Q(f)  & U(f)-iV(f)\\
U(f) + iV(f) & I(f) - Q(f)
\end{pmatrix},
\end{equation}
\end{widetext}
with $\delta(\cdot)$ denotes Dirac delta functions, and
$\delta^2(\hat{k}-\hat{k}') = \delta(\phi-\phi') \delta(\cos\theta-\cos\theta')$.
In Eq.\eqref{eq:stocks}, $\langle \cdots \rangle$ denotes the ensemble average,  $I$, $Q$, $U$, and $V$ are the Stokes parameters. Specifically, $I(f)$ represents the total intensity of the background, $V(f)$ characterizes the asymmetry in the intensities of right- and left-hand circular polarization, corresponding to the net circular polarization, while $Q(f)$ and $U(f)$ describe the linear polarization. For an isotropic stochastic background, the contributions from $Q$ and $U$ vanish, as pointed out in previous studies~\citep{Seto2006,Seto2007a}. Therefore, we omit these two parameters in our subsequent analysis. We ignore the effects of the shot noise from a finite number of GW sources which will cause $Q$ and $U$ to be non-zero~\citep{Belgacem:2024ohp}.

It is more convenient to introduce the circular polarization tensors:
\begin{equation}
e^R_{ab} = \frac{e^+_{ab} + i e^\times_{ab}}{\sqrt{2}}, \quad e^L_{ab} = \frac{e^+_{ab} - i e^\times_{ab}}{\sqrt{2}},
\end{equation}
and the corresponding strains:
\begin{equation}
h_R = \frac{h_+ - i h_\times}{\sqrt{2}}, \quad h_L = \frac{h_+ + i h_\times}{\sqrt{2}}.
\end{equation}
Eq.~\eqref{eq:stocks} then becomes:
\begin{equation}
\label{eq:circular}
\langle\tilde{h}_\lambda(f, \hat{k})\tilde{h}^*_{\lambda'}(f', \hat{k}')\rangle = \frac{1}{2}\delta(f-f') \frac{\delta^2(\hat{k}-\hat{k}')}{4\pi} \hat{\delta}_{\lambda \lambda'} S_\lambda(f),
\end{equation}
with $\lambda = R, L$.  Here, $\hat{\delta}_{\lambda\lambda'}$ is the Kronecker delta, and subscripts are used to distinguish it from the Dirac delta. $S_{\lambda}(f)$ is the spectral density of the right- or left-hand circular polarization. These spectral densities are related to the Stokes parameters through:
\begin{align}
\label{eq:LR}
I(f) &= \frac{1}{2} [S_R(f) + S_L(f)], \\
V(f) &= \frac{1}{2} [S_R(f) - S_L(f)].
\end{align}

In the frequency domain, the detector response to the SGWB signal is \citep{Maggiore2007}:
\begin{equation}
\label{eq:signal}
\tilde{s}(f) = \sum_{\lambda=R, L} \int d^2\hat{k} \, F^\lambda(f, \hat{k}) \, \tilde{h}_\lambda (f,\hat{k}) \, e^{-i 2\pi f\hat{k}\cdot\vec{r}/c},
\end{equation}
where $F^\lambda(f, \hat{k})$ is the detector response function to a GW signal of polarization $\lambda$ from direction $\hat{k}$, and $\vec{r}$ is the position vector of the detector.

For the correlation between two detectors $i$ and $j$, we have \citep{Seto2007a,Seto2008}:
\begin{equation}
\label{eq:correlation}
\langle \tilde{s}_i(f) \tilde{s}^*_j(f')\rangle = \frac{1}{2}\delta(f-f') \left[\Gamma^I_{ij}(f)I(f) + \Gamma^V_{ij}(f)V(f) \right],
\end{equation}
where
\begin{equation}
\label{eq:overlap}
\begin{aligned}
\Gamma^I_{ij}(f) &\equiv \int \frac{d^2\hat{k}}{4\pi} \left(F^+_i F^{+*}_j + F^\times_i F^{\times*}_j\right) e^{-i 2\pi f\hat{k}\cdot \Delta\vec{r}/c},\\
\Gamma^V_{ij}(f) &\equiv -{\rm i} \int \frac{d^2\hat{k}}{4\pi} \left(F^+_i F^{\times*}_j - F^\times_i F^{+*}_j\right) e^{-i 2\pi f\hat{k}\cdot \Delta\vec{r}/c}.
\end{aligned}
\end{equation}
Here, $\Delta \vec{r} \equiv \vec{r}_i - \vec{r}_j$ is the separation vector between the two detectors. $\Gamma^I_{ij}$ and $\Gamma^V_{ij}$ are the overlap reduction functions (ORFs) for the total background and the circular polarization components of the isotropic background, respectively, describing the correlation between the responses of the two detectors to these components.

\section{Sensitivity of LISA-TAIJI network for polarized SGWB} \label{sec:result}

\begin{figure*}[htp]
 \includegraphics[width=0.45\textwidth]{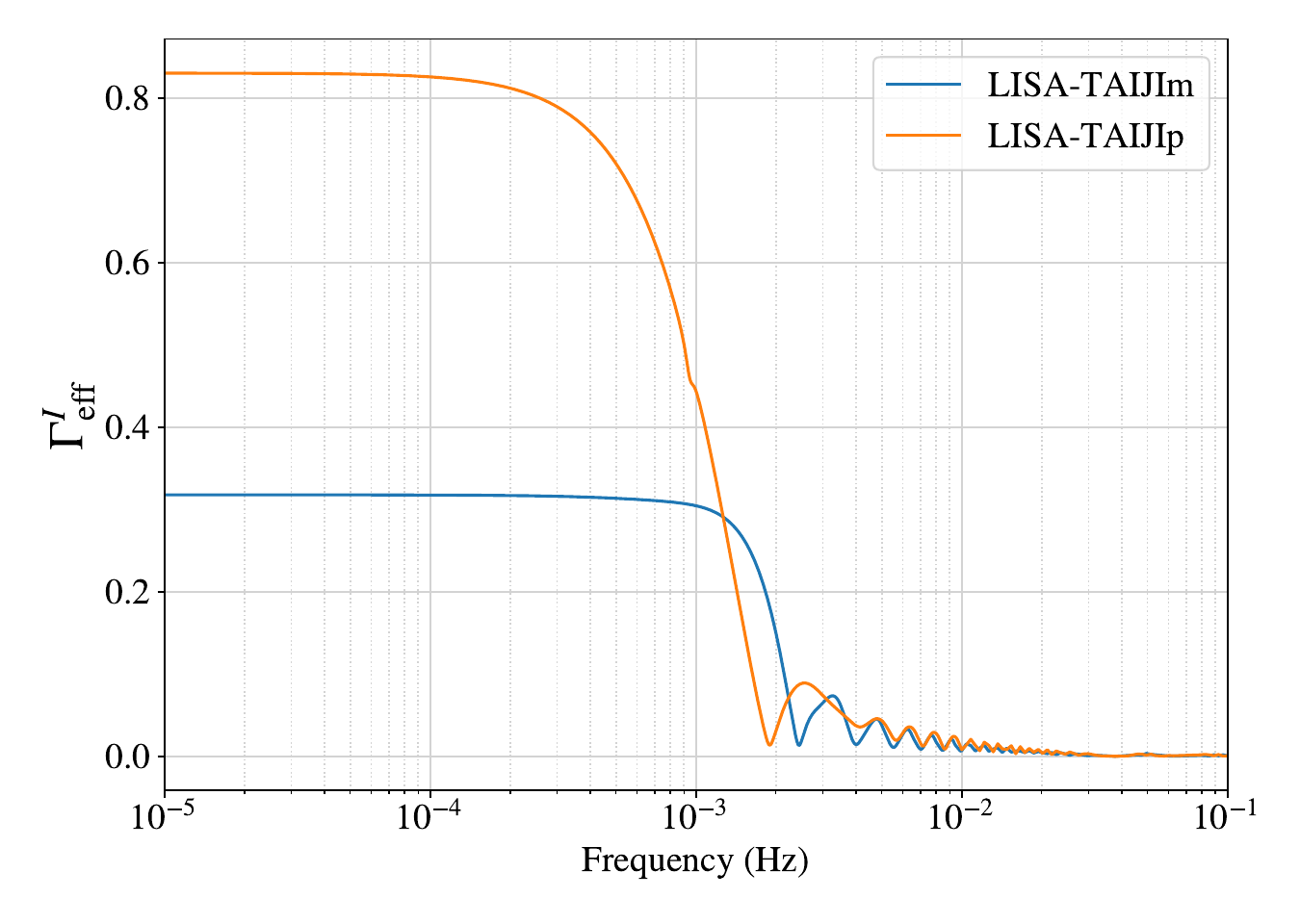} 
 \includegraphics[width=0.45\textwidth]{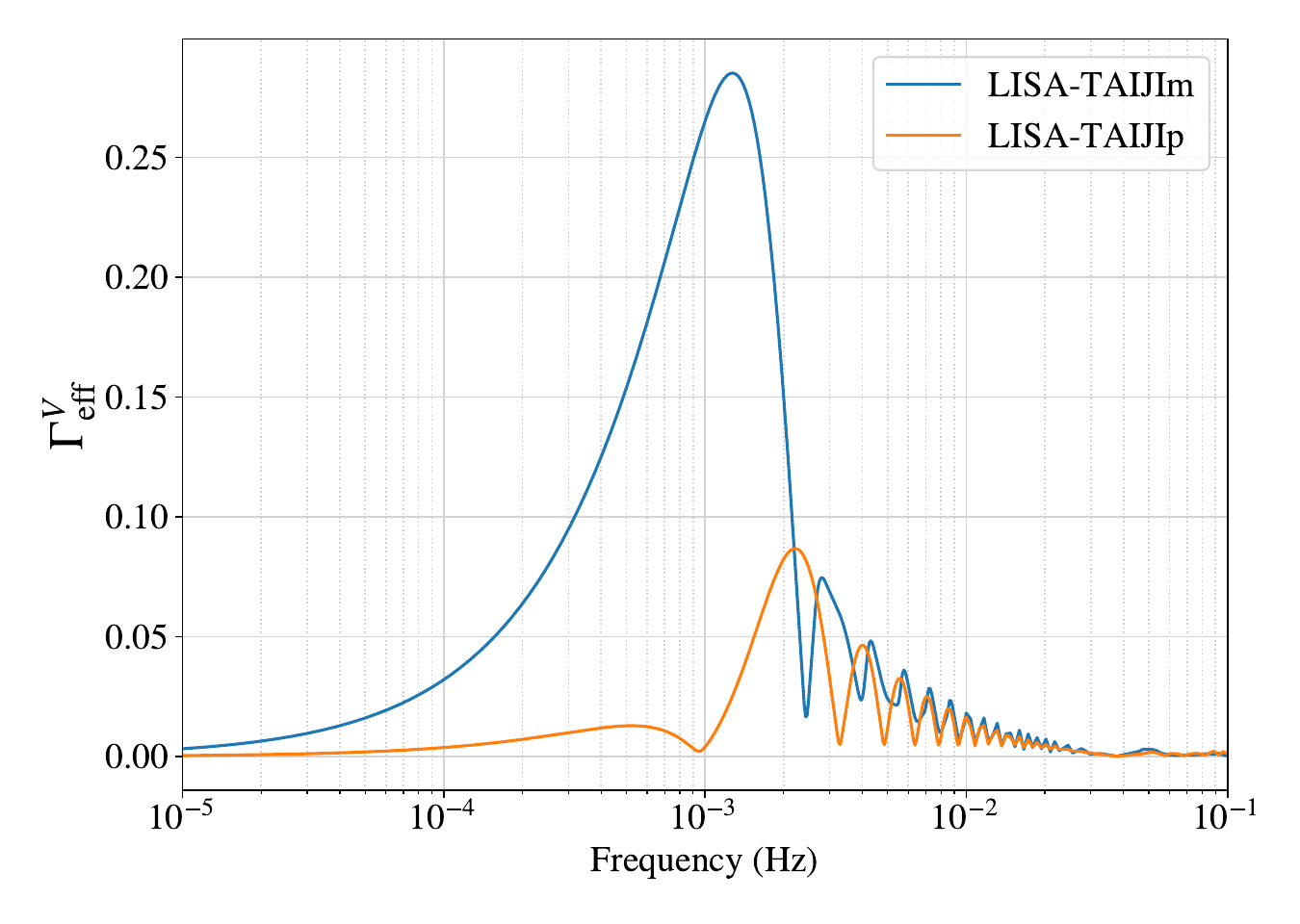} 
 \includegraphics[width=0.45\textwidth]{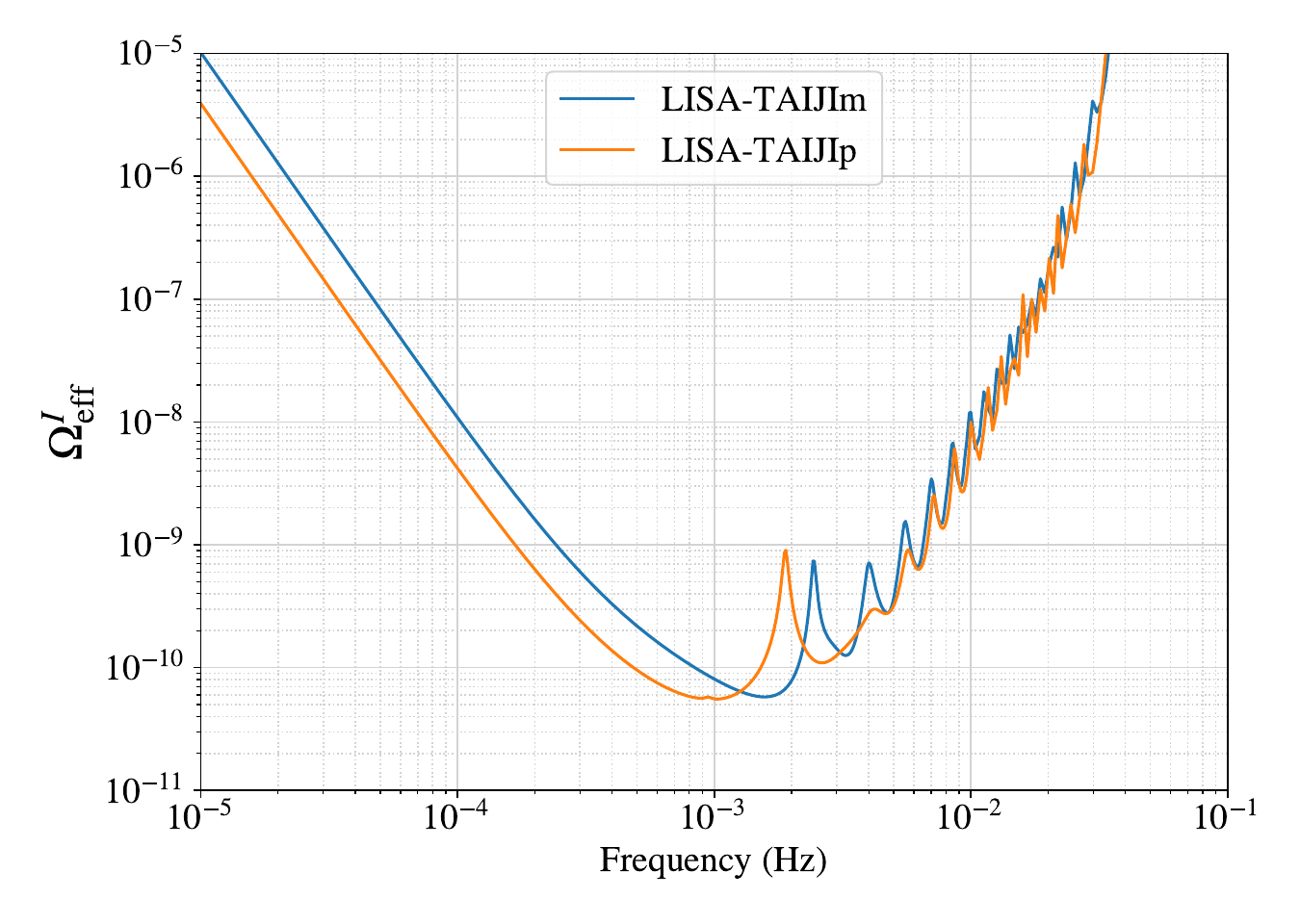} 
 \includegraphics[width=0.45\textwidth]{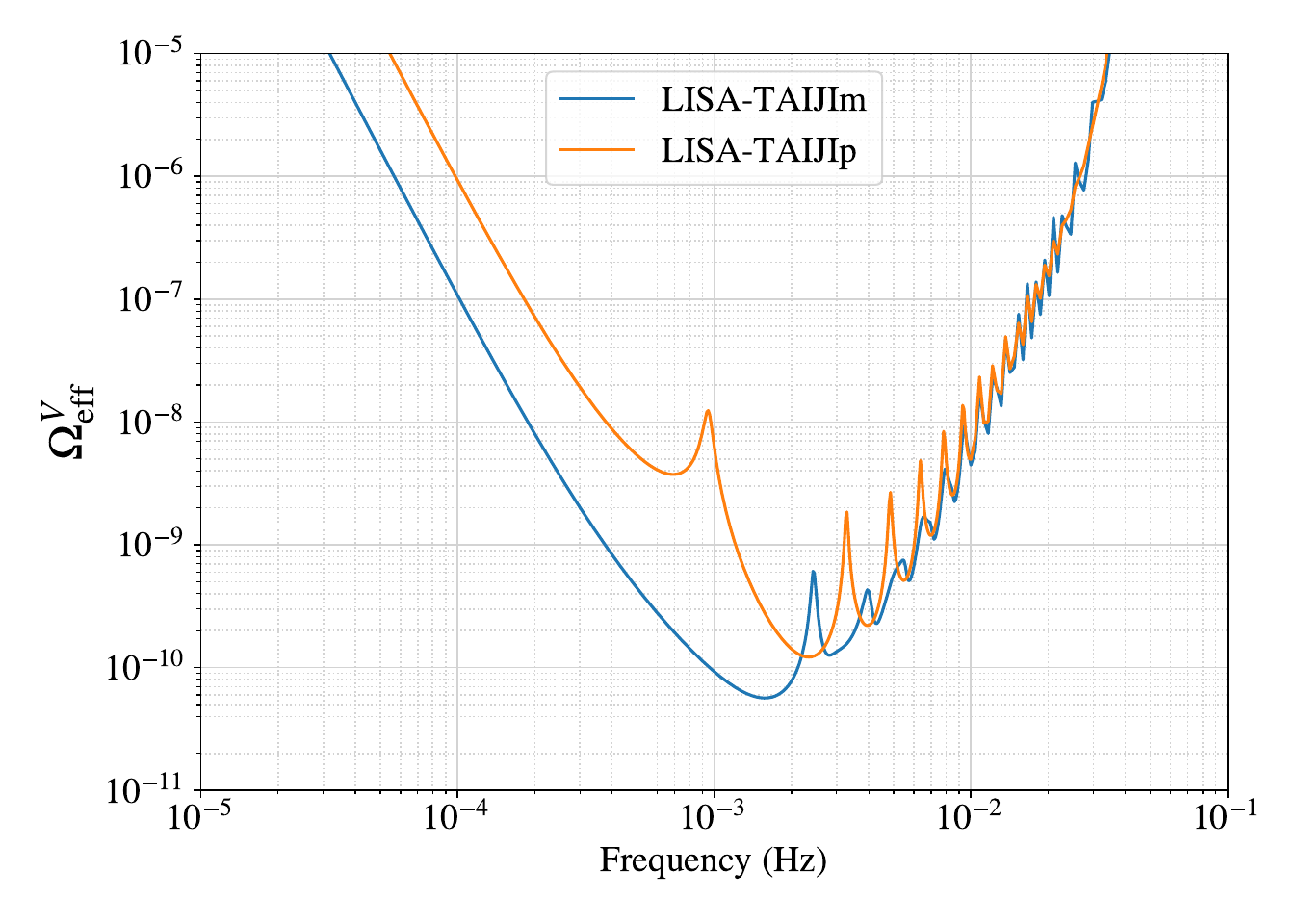}
 \caption{The effective overlap reduction functions (top) and average sensitivities (bottom) of the LISA-TAIJI networks for the $I$ and $V$ components. Results for LISA-TAIJIp and LISA-TAIJIm are shown in orange and blue, respectively.
}
 \label{fig:ORF}
\end{figure*}

The observational data are modeled as the sum of the detector response $\tilde{s}(f)$ and noise $\tilde{n}(f)$:
\begin{equation}
    \tilde{d}(f) = \tilde{s}(f) + \tilde{n}(f).
\end{equation}
Assuming stationary noises, the noises in the channels of different detectors are considered uncorrelated. Therefore,
\begin{equation}
    \langle \tilde{n}_i(f)\tilde{n}_j^*(f') \rangle = \frac{1}{2}\hat{\delta}_{ij}\delta(f-f')N_i(f),
\end{equation}
where $N_i(f)$ represents the one-sided noise power spectral density of detector (channel) $i$.

The cross-correlations of the observed data from two channels of the detectors $i$ and $j$ in the frequency domain are defined as:
\begin{equation}
\mathcal{C}_{ij}(f) \equiv \frac{1}{T_{\rm obs}} \tilde{d}_i(f) \tilde{d}^*_j(f),
\end{equation}
where $T_{\rm obs}$ is the observation time. Since the signal and noise are uncorrelated, and the noises of different detectors are also uncorrelated, the expectation value of the cross-correlation simplifies to $\langle \mathcal{C}_{ij} \rangle = \frac{1}{T_{\rm obs}}\langle \tilde{s}_i(f)\tilde{s}^*_j(f) \rangle$.
As shown in Eq.~\eqref{eq:correlation}, for finite observation time $T_{\rm obs}$, when
$f=f'$, the Dirac delta function turns out to be $T_{\rm obs}$,
such that
$\langle \tilde{s}_i(f) \tilde{s}^*_j(f)\rangle = \frac{T_{\rm obs}}{2}\left[\Gamma^I_{ij}(f)I(f) + \Gamma^V_{ij}(f)V(f) \right]$. Considering these results, we obtain:
\begin{equation}
    \langle \mathcal{C}_{ij} \rangle = \frac{1}{2} \left[\Gamma^I_{ij}(f) I(f) + \Gamma^V_{ij}(f) V(f)\right].
\end{equation}

Under the weak signal approximation, the variance of the cross-correlation is 
(see Eq.(6.24) in \cite{Romano2017}):
\begin{align}
\label{eq:variance}
\sigma^2_{{ij}}=    &\langle\mathcal{C}_{ij}(f)\mathcal{C}^*_{kl}(f')\rangle - \langle\mathcal{C}_{ij}(f)\rangle\langle\mathcal{C}^*_{kl}(f')\rangle \nonumber\\
\approx  & \frac{1}{4} \frac{1}{T_\mathrm{obs}} N_i(f) N_j(f') \hat{\delta}_{ik}\hat{\delta}_{jl} \delta(f-f').
\end{align}
The square of the signal-to-noise ratio (SNR) for the signal in a small frequency band $\delta f$ is given by $T_{\rm obs} {\delta f} \frac{\langle \mathcal{C}_{ij} \rangle^2}{\sigma^2_{ij}}$ (see Eq.(8.14) in \cite{Romano2017}). 
Summing over all frequency bands gives:
\begin{equation}
\begin{aligned}
    {\rho}_{ij}^2 &=  T_{\rm obs} \int^{\infty}_{-\infty} \mathrm{d}f \frac{\langle \mathcal{C}_{ij} \rangle^2}{\sigma^2_{ij}} \\
    &= T_\mathrm{obs} \left[ \int^{\infty}_{-\infty} \mathrm{d}f \frac{\left[ \Gamma^I_{ij}(f) I(f) + \Gamma^V_{ij}(f) V(f) \right]^2}{N_i(f) N_j(f)} \right].
    \label{eq:SNRij}
\end{aligned}
\end{equation}
This formula is the same as the maximum SNR obtained with the optimal filter method \citep{Seto2008, Orlando2021}.

For strong GW signals, the $N_i(f) N_j(f)$ term in the denominator of Eq.~\eqref{eq:SNRij} should be replaced by \citep{Cornish:2001qi, Cornish2002}:
\begin{equation}
\begin{aligned}
 M_{ij}(f) = \left( N_{i}  + \mathcal{R}_i I \right) \left( N_{j} + \mathcal{R}_j I \right) + \left( \Gamma^{I}_{ij} I + \Gamma^{V}_{ij} V \right)^2 .
 \label{eq:M_noApprox}
\end{aligned}
\end{equation}
where $\mathcal{R}_i$ and $\mathcal{R}_j$ are the averaged responses of a TDI channel over all sky directions, defined as:
\begin{equation}
\label{Rtdi}
\mathcal{R}_i(f) = \int \frac{\mathrm{d}^2 \hat{k}}{4\pi} \left[\left|F_i^+(f, \hat{k})\right|^2 + \left|F_i^\times(f, \hat{k})\right|^2\right],
\end{equation}
where $F_i(f, \hat{k})$ represents the TDI response to a GW signal from direction $\hat{k}$.

\subsection{Effective overlap reduction functions}

With the LISA-TAIJI network, we have four independent channel pairs, denoted as $\kappa \in \{ A_L{\text -}A_T$, $A_L{\text -}E_T$, $E_L{\text -}A_T$, $E_L{\text -}E_T \}$. The total SNR for the stochastic background is then given by:
\begin{equation}
    \label{eq:SNR_total}
    \rho^2 = T_\mathrm{obs}  \sum_{\kappa} \int^{\infty }_{-\infty} \mathrm{d} f  \frac{\left[ \Gamma^I_{\kappa} (f) I(f) + \Gamma^V_{\kappa} (f) V(f) \right]^2}{N^2_{\kappa}(f)},
\end{equation}
where $N_{\kappa}(f) \equiv \sqrt{N_L(f) N_T(f)}$, with $N_L(f)$ and $N_T(f)$ being the noise spectra of the $A$ or $E$ channels for the LISA and TAIJI detectors, respectively.

Eq.~\eqref{eq:SNR_total} provides the SNR for the SGWB in general, without distinguishing between the $I$ and $V$ components. To detect circular polarization, it is necessary to separate the $I$ and $V$ components from the cross-correlation $\mathcal{C}_{\kappa}$ of the observed data. This separation requires multiple detector (channel) pairs, as described in \citep{Seto2007, Seto2008}. We provide a brief introduction to this approach below.

Consider a small frequency band $\delta f$, where $\Gamma^I$, $\Gamma^V$, $I$, and $V$ can be treated as constants. From Eq.~\eqref{eq:variance}, the variance of $\mathcal{C}_{\kappa}(f)$ for each channel pair is $\frac{1}{4} N^2_{\kappa}(f)$, while the covariance between different channel pairs is zero. Assuming Gaussian noise, the likelihood of the signal model in this frequency band is given by \citep{Seto2008,Romano2017}:
\begin{equation}
    p(\mathcal{C}|I,V) \propto \exp\left\{ -\frac{T_{\rm obs} \delta f }{2}  \sum_{\kappa} \frac{\left[2\mathcal{C}_{\kappa} - (\Gamma^I_{\kappa} I + \Gamma^V_{\kappa} V)\right]^2}{N^2_{\kappa}(f)} \right\}.
\end{equation}

The Fisher matrix for the parameters $I$ and $V$ is then:
\begin{equation}
  \begin{aligned}
    \mathcal{F} &= - \left\langle \frac{\partial^2 \ln p(\mathcal{C}|\mathcal{S})}{\partial\theta\partial\theta} \right\rangle \\
        &= T_{\rm obs} ~ \delta f \begin{pmatrix}
        \sum_{\kappa} \frac{(\Gamma^I_{\kappa})^2}{N^2_{\kappa}} &  \sum_{\kappa} \frac{\Gamma^I_{\kappa}\Gamma^V_{\kappa}}{N^2_{\kappa}}\\
        \sum_{\kappa} \frac{\Gamma^V_{\kappa}\Gamma^I_{\kappa}}{N^2_{\kappa}} & \sum_{\kappa} \frac{(\Gamma^V_{\kappa})^2}{N^2_{\kappa}}
        \end{pmatrix}.
  \end{aligned}
\end{equation}

The covariance matrix of the signal parameters is the inverse of the Fisher matrix. Therefore, the variances of the estimators for $I$ and $V$ are:
\begin{equation}
    \sigma^2_{\hat{I}}(f) = \frac{\mathcal{F}_{22}}{|\mathcal{F}|}, ~~~ \sigma^2_{\hat{V}}(f) = \frac{\mathcal{F}_{11}}{|\mathcal{F}|}.
\end{equation}
The SNRs for the $I$ and $V$ components in the frequency band are then:
\begin{equation}
    \begin{aligned}
      \frac{\hat{I}^2(f)}{\sigma^2_{\hat{I}}(f)} &= T_{\rm obs} \delta f ~I^2  \left[\sum_{\kappa} \frac{(\Gamma^I_{\kappa})^2}{N^2_{\kappa}} - \frac{\left(\sum_{\kappa} \frac{\Gamma^I_{\kappa}\Gamma^V_{\kappa}}{N^2_{\kappa}}\right)^2  }{ \sum_{\kappa} \frac{(\Gamma^V_{\kappa})^2}{N^2_{\kappa}} } \right],\\
      \frac{\hat{V}^2(f)}{\sigma^2_{\hat{V}}(f)} &= T_{\rm obs} \delta f~V^2 \left[\sum_{\kappa} \frac{(\Gamma^V_{\kappa})^2}{N^2_{\kappa}} - \frac{\left(\sum_{\kappa} \frac{\Gamma^I_{\kappa}\Gamma^V_{\kappa}}{N^2_{\kappa}}\right)^2  }{ \sum_{\kappa} \frac{(\Gamma^I_{\kappa})^2}{N^2_{\kappa}} } \right].
    \end{aligned}
\end{equation}

Summing over all frequency bands and noting that the noise spectra $N_{\kappa}$ are the same for all four channel pairs, we obtain:
\begin{equation}
    \begin{aligned}
      \label{eq:SNR}
      \rho_I^2 &= T_{\rm obs} \int \mathrm{d}f \left[\sum_{\kappa} (\Gamma^I_{\kappa})^2 - \frac{\big(\sum_{\kappa} \Gamma^I_{\kappa}\Gamma^V_{\kappa}\big)^2}{\sum_{\kappa} (\Gamma^V_{\kappa})^2}\right] \frac{I^2}{N^2},\\
      \rho_V^2 &= T_{\rm obs} \int \mathrm{d}f \left[\sum_{\kappa} (\Gamma^V_{\kappa})^2 - \frac{\big(\sum_{\kappa} \Gamma^I_{\kappa}\Gamma^V_{\kappa}\big)^2}{\sum_{\kappa} (\Gamma^I_{\kappa})^2}\right] \frac{V^2}{N^2} .     
    \end{aligned}
  \end{equation}
We now introduce the effective ORFs following \citep{Seto2008}:
\begin{equation}
  \begin{aligned}
    \label{eq:overlap_eff}
    \Gamma^{I}_{\rm eff} = \sqrt{\sum_{\kappa} (\Gamma^I_{\kappa})^2 - \frac{\big(\sum_{\kappa} \Gamma^I_{\kappa}\Gamma^V_{\kappa}\big)^2}{\sum_{\kappa} (\Gamma^V_{\kappa})^2}}\,,\\
    \Gamma^{V}_{\rm eff} = \sqrt{\sum_{\kappa} (\Gamma^V_{\kappa})^2 - \frac{\big(\sum_{\kappa} \Gamma^I_{\kappa}\Gamma^V_{\kappa}\big)^2}{\sum_{\kappa} (\Gamma^I_{\kappa})^2}} \,.  
  \end{aligned}
\end{equation}
The optimal SNR is thus fully determined by the effective ORFs $\Gamma^{I}_{\rm eff}$ and $\Gamma^{V}_{\rm eff}$, as well as the spectrum shapes $I(f)$ and $V(f)$.

The plots of $\Gamma^I_{\rm eff}$ and $\Gamma^V_{\rm eff}$ for the two network configurations are shown in the upper row of Fig.~\ref{fig:ORF}, where they have been normalized to unity for the co-located and co-aligned configuration. The orange and blue lines correspond to the LISA-TAIJIp and LISA-TAIJIm configurations, respectively.
For the intensity $I$ component (left column), the LISA-TAIJIm configuration exhibits lower correlation at lower frequencies compared to LISA-TAIJIp. This is due to the more misaligned orientation of the two detectors' constellations. The correlation surpasses that of LISA-TAIJIp around 2 mHz, which corresponds to the most sensitive frequency band for these detectors. At higher frequencies, the ORFs for both configurations are nearly identical. These trends are reflected in the sensitivities of the joint observations, as shown in the lower left plot of Fig.~\ref{fig:ORF}.

For the $V$ component, the ORFs for both configurations are generally low in the low-frequency band as shown in the upper right plot of Fig.~\ref{fig:ORF}. LISA-TAIJIm consistently outperforms LISA-TAIJIp across nearly all frequencies, apart from minor dips caused by the cross-term $\sum_{\kappa} \Gamma^I_{\kappa}\Gamma^V_{\kappa}$ in $\Gamma^V_{\rm eff}$. Notably, in the low-frequency limit, the ORF for LISA-TAIJIp becomes negligible, which explains why the SNR for the $V$ component remains almost constant as the lower frequency limit decreases below 2 mHz \citep{Seto2020a}. In contrast, LISA-TAIJIm shows significantly better correlation in the low-frequency regime with the superior ORF. Consequently, the sensitivity of LISA-TAIJIm surpasses that of LISA-TAIJIp, as shown in the lower right panel of Fig.~\ref{fig:ORF}.

\subsection{Power-law integrated sensitivity curve }

The power-law integrated (PLI) sensitivity curve is a common method for representing the sensitivity of detectors to SGWB with a power-law spectrum \citep{Thrane2013}. It can be obtained using the SNR provided in Eq.~\eqref{eq:SNR}, along with the effective ORFs defined above. Following the convention, we use the fractional energy density spectrum $\Omega(f)$ in the sensitivity curve, which is related to power spectrum density $I(f), V(f)$ by:
\begin{equation} \label{eq:Omega}
  \Omega^{I}(f) = \frac{4\pi^2 f^3}{3H_0^2} I(f),\qquad \Omega^{V}(f) = \frac{4\pi^2 f^3}{3H_0^2} V(f).
\end{equation}

For a specific spectrum index $\alpha$, the fractional energy density spectrum $\Omega(f) = \Omega_\alpha \left(\frac{f}{f_c}\right)^\alpha$. Given the detection threshold of the SNR, $\rho_{\rm thr}$, the minimal detectable SGWB spectrum can be obtained using Eq.~\eqref{eq:SNR} and Eq.~\eqref{eq:Omega}:
\begin{equation}
  \Omega^{\{I,V\}}_{\alpha_{\rm thr}} =  \frac{{\rho}_{\rm thr}}{\sqrt{T_{\rm obs}}} \frac{4\pi^2}{3H_0^2} \left[\int \mathrm{d}f \frac{\Gamma^{\{I,V\}^2}_{\rm eff}(f) f^{2\alpha-6}}{N^2(f)}\right]^{-\frac{1}{2}} f_c^\alpha\,.
  \label{eq:Omega_limit}
\end{equation}
Here, we choose $T_{\rm obs} = 3~\rm year$, and set ${\rho}_{\rm thr} = 10$.

\begin{figure*}[htp]
 \includegraphics[width=0.45\textwidth]{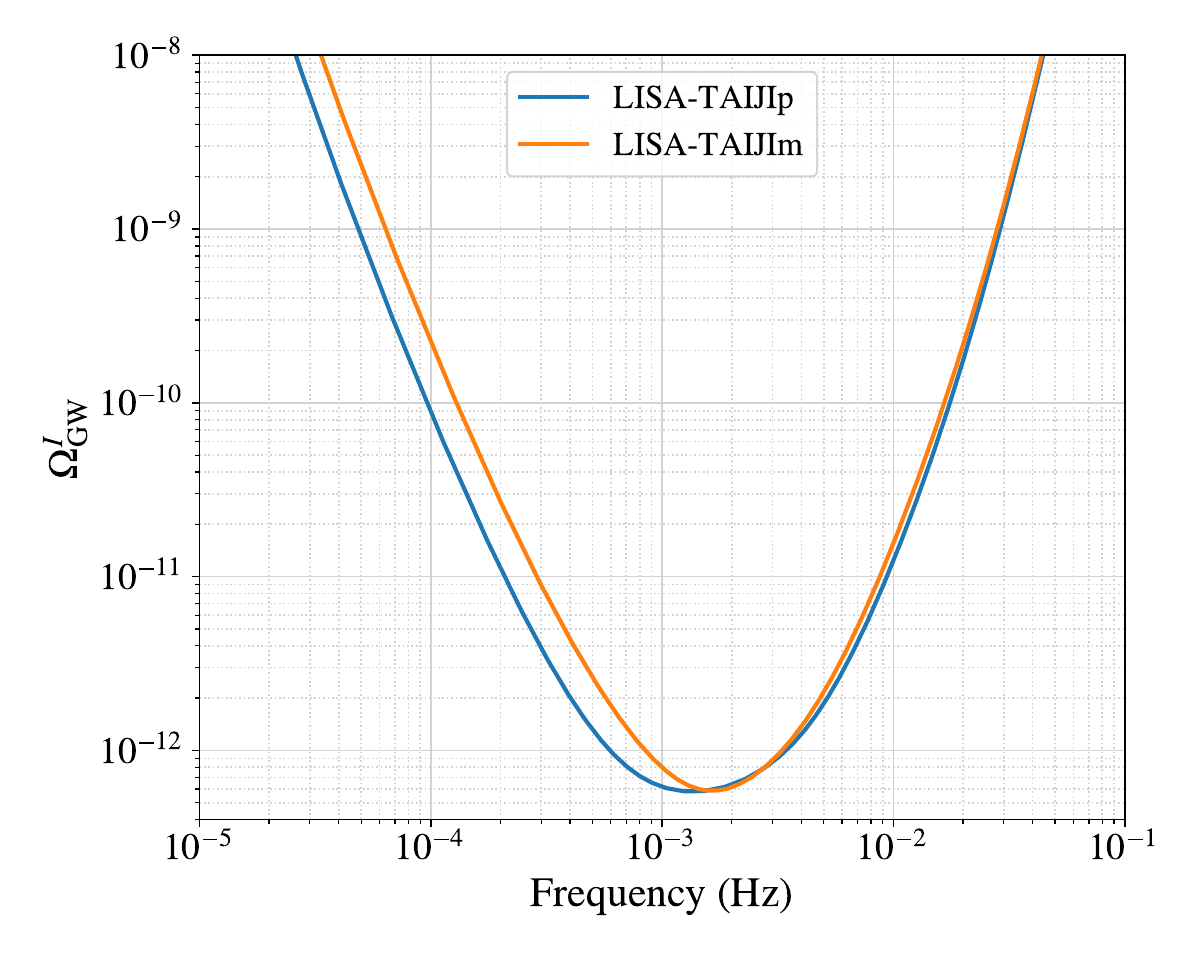}
 \includegraphics[width=0.45\textwidth]{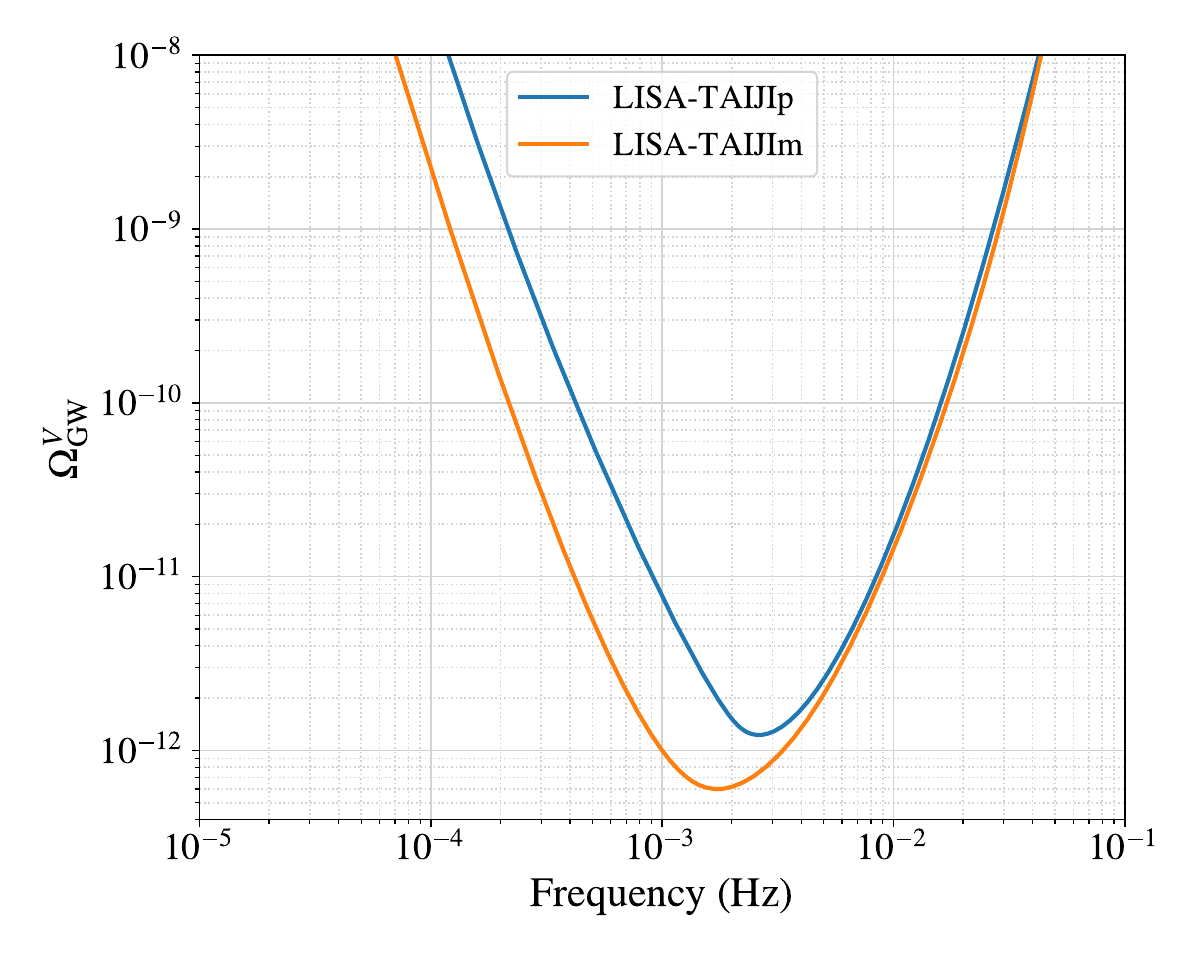}
 \caption{ The power-law integrated sensitivity curves for the $I$ and $V$ components of different LISA-TAIJI networks. The LISA-TAIJIp and LISA-TAIJIm configurations are represented by the blue and orange lines, respectively. The SNR threshold is set to 10, with an observation time of $T_{\rm obs} = 3~\rm years$.}
 \label{fig:PLI}
\end{figure*}

The PLI sensitivity is defined as the (upper) envelope of the fractional energy density spectrum for all possible spectrum indices.
\begin{equation}
  \Omega^{\{I,V\}}_{\rm PLI}(f) = \max_{\alpha}\left[\Omega^{\{I,V\}}_{\alpha_{\rm thr}} \left(\frac{f}{f_c}\right)^\alpha\right].
\end{equation}
With the effective ORFs given in Fig.~\ref{fig:ORF}, we can compute the PLI sensitivity for different networks, as shown in Fig.~\ref{fig:PLI}.
The left panel shows the PLI sensitivity for the overall intensity $I$, with the blue and orange lines representing the LISA-TAIJIp and LISA-TAIJIm configurations, respectively. In the low-frequency range, the sensitivity of LISA-TAIJIm is lower compared to the default configuration, while the sensitivities become comparable around 2 mHz and at higher frequencies. These trends are consistent with the ORF behavior discussed earlier.
The right panel presents the sensitivity for the circular polarization component $V$. At low frequencies, the LISA-TAIJIm configuration exhibits a significant improvement in sensitivity compared to the default configuration, nearly an order of magnitude, as expected from the ORF results. At high frequencies, however, the sensitivities of both configurations are nearly identical. Notably, with LISA-TAIJIm, we can achieve comparable sensitivity for both the $I$ and $V$ components in the most sensitive frequency range, around 2 mHz.

\section{Determining the parity of SGWB with LISA-TAIJI network}
\label{sec:parameters} 

To further quantify the performance of different LISA-TAIJI networks, we select three energy density spectrum models, compare the SNR for the parity of the SGWB, and examine the parameter constraints based on Fisher information forecasts.

\subsection{SGWB energy spectrum models}

Here, we introduce the three models of the energy density spectrum used in this study. For a comprehensive overview of the various cosmological and astrophysical processes contributing to the SGWB, see the reviews in \cite{Binetruy:2012ze,Caprini:2018mtu,Kuroyanagi:2018csn,LISACosmologyWorkingGroup:2022jok}.

1) Power-Law Model~\cite{Turner:1993vb,Giovannini:1998bp,Peebles:1998qn,Giovannini:1999bh,Giovannini:1999qj,Cai:2019cdl,Auclair:2019wcv,Tashiro:2003qp,Giovannini:2008tm,Domenech:2020kqm}:
\begin{equation}
\Omega_\mathrm{PL} = \Omega_{1} \left( \frac{f}{f_c} \right)^{\alpha_1}. \label{eq:PL_signal}
\end{equation}
The fiducial signal is defined with $\Omega_1 = 4.446 \times 10^{-12}$ and $\alpha_1 = 2/3$ at the reference frequency $f_c = 1 \ \mathrm{mHz}$ \cite{LIGOScientific:2019vic}. The power-law spectrum is very common in cosmological processes. Parity violation can arise in scenarios involving pseudo-scalar inflatons or modified gravity~\cite{Alexander:2004us,Takahashi:2009wc,Magueijo:2010ba,Sorbo:2011rz}. Measurements of such parity violations with ground-based GW detectors have been explored in \cite{Seto:2007tn,Crowder:2012ik}.

2) Single Peak Model~\cite{Kohri:2018awv,Inomata:2019ivs,Cai:2019amo,White:2021hwi,Lozanov:2022yoy}:
\begin{equation}
 \Omega_{\rm SP} = \Omega_2 \exp\left[ - \frac{ \left( \log_{10} ( f /f_c ) \right)^2}{ \Delta^2_2 } \right]. \label{eq:SP_signal}
\end{equation}
The typical parameters are set to  $\Delta_2 = 0.2$, $\Omega_2 = 1 \times 10^{-11}$, and $f_c = 3 \ \mathrm{mHz}$ \cite{Caprini:2019pxz,Flauger:2020qyi}. Single-peak SGWB spectrum can also arise in cosmological processes, with parity-violating GWs potentially generated during inflation through mechanisms such as dynamical Chern-Simons gravity~\cite{Fu:2020tlw,Peng:2022ttg}.

3) Broken Power-Law Model~\cite{Vilenkin:1981zs,Kosowsky:1992rz,Kamionkowski:1993fg,Gleiser:1998na,Grojean:2006bp,Hindmarsh:2013xza,Caprini:2015zlo,Saikawa:2017hiv,Caprini:2019egz}:
\begin{equation}
 \Omega_{\rm BPL} = \Omega_3 \left( \frac{f}{ f_c } \right)^{\alpha_2} \left[ 1 + 0.75 \left( \frac{f}{ f_c } \right)^\Delta \right]^{  (\alpha_3 - \alpha_2 ) / \Delta }. \label{eq:BPL_signal}
\end{equation}
The fiducial parameters are assumed to be $\alpha_2 = 3$, $\alpha_3 = -4$, $\Delta = 2$, with an amplitude $\Omega_3 = 1 \times 10^{-9}$ and reference frequency $f_c = 10 \ \mathrm{mHz}$, see e.g.~\cite{Hindmarsh:2013xza,Caprini:2015zlo,Caprini:2019egz,Martinovic:2020hru}. 
The broken power-law spectrum is commonly produced during first-order phase transitions, see e.g.~\cite{RoperPol2020}. A prominent source of parity-violating backgrounds with this spectrum is axion-like particles, which are also viable candidates for dark matter~\cite{Machado:2018nqk,Machado:2019xuc,Ding:2024,Xu:2024kwy}.

Finally, to describe the circularly polarized components, we introduce the polarization degree parameter:
\begin{eqnarray}
    \Pi(f) = V(f)/ I(f),
\end{eqnarray}
which quantifies the strength of parity violation. For simplicity, we take $\Pi(f)$ as constant over all frequency ranges in our analysis. By definition, $\Pi \in [-1, 1]$, where $\Pi = 0$ corresponds to an unpolarized background, while $\Pi = \pm 1$ corresponds to fully left-handed or right-handed circular polarization, respectively.

\begin{figure*}[htp]
 \includegraphics[width=0.45\textwidth]{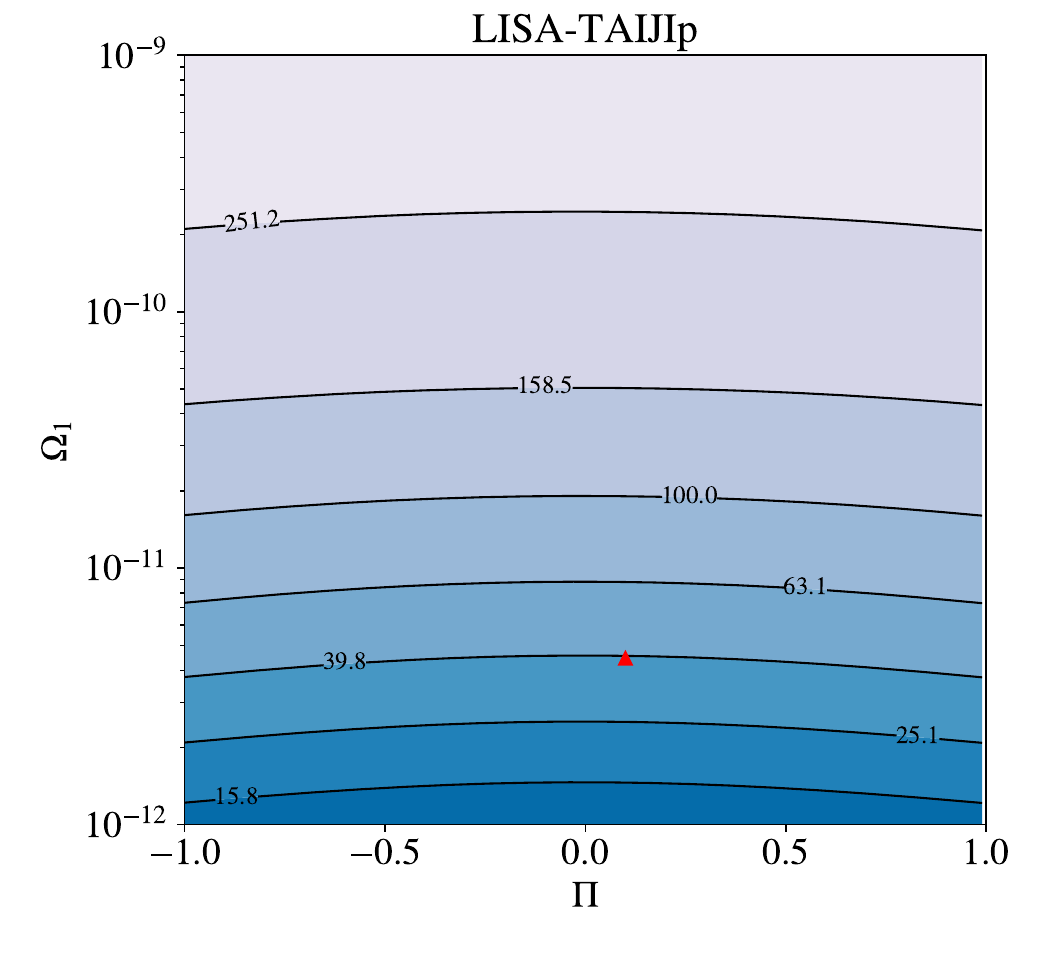}
 \includegraphics[width=0.45\textwidth]{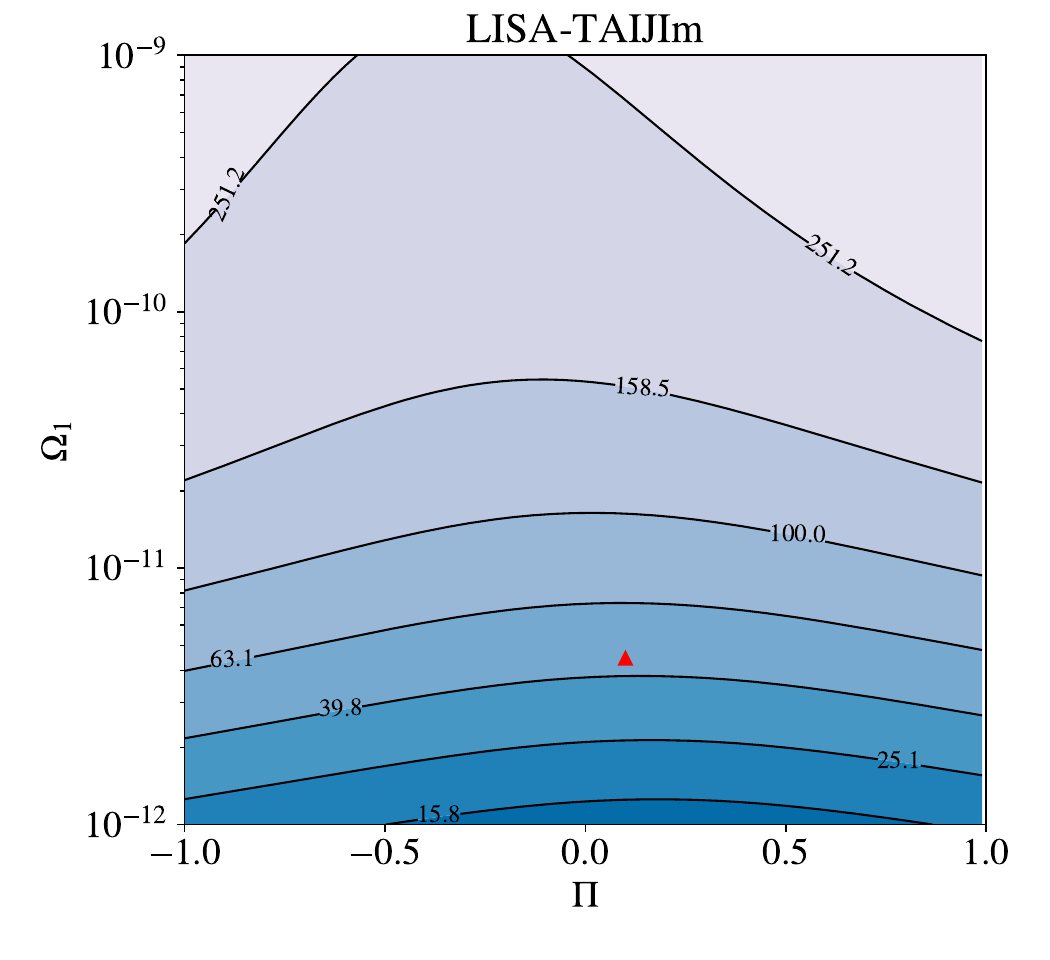}
 \includegraphics[width=0.45\textwidth]{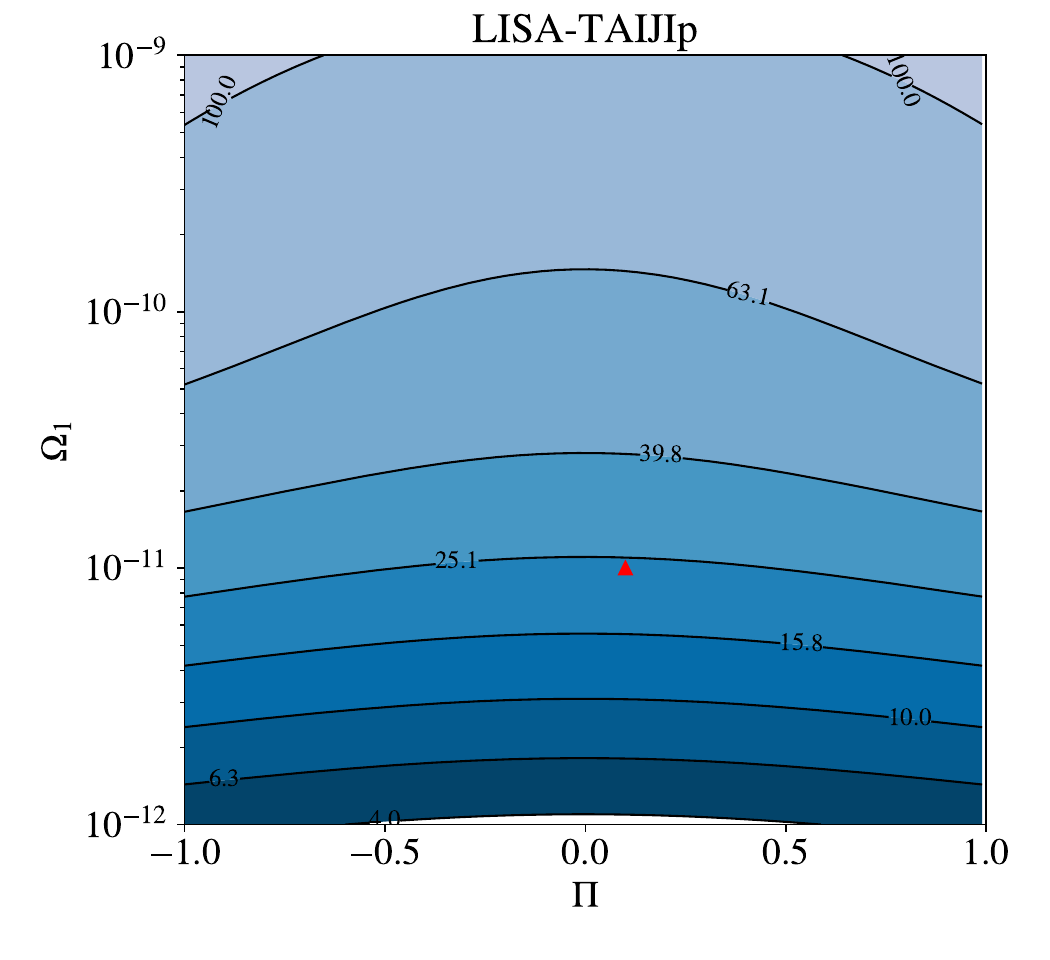}
 \includegraphics[width=0.45\textwidth]{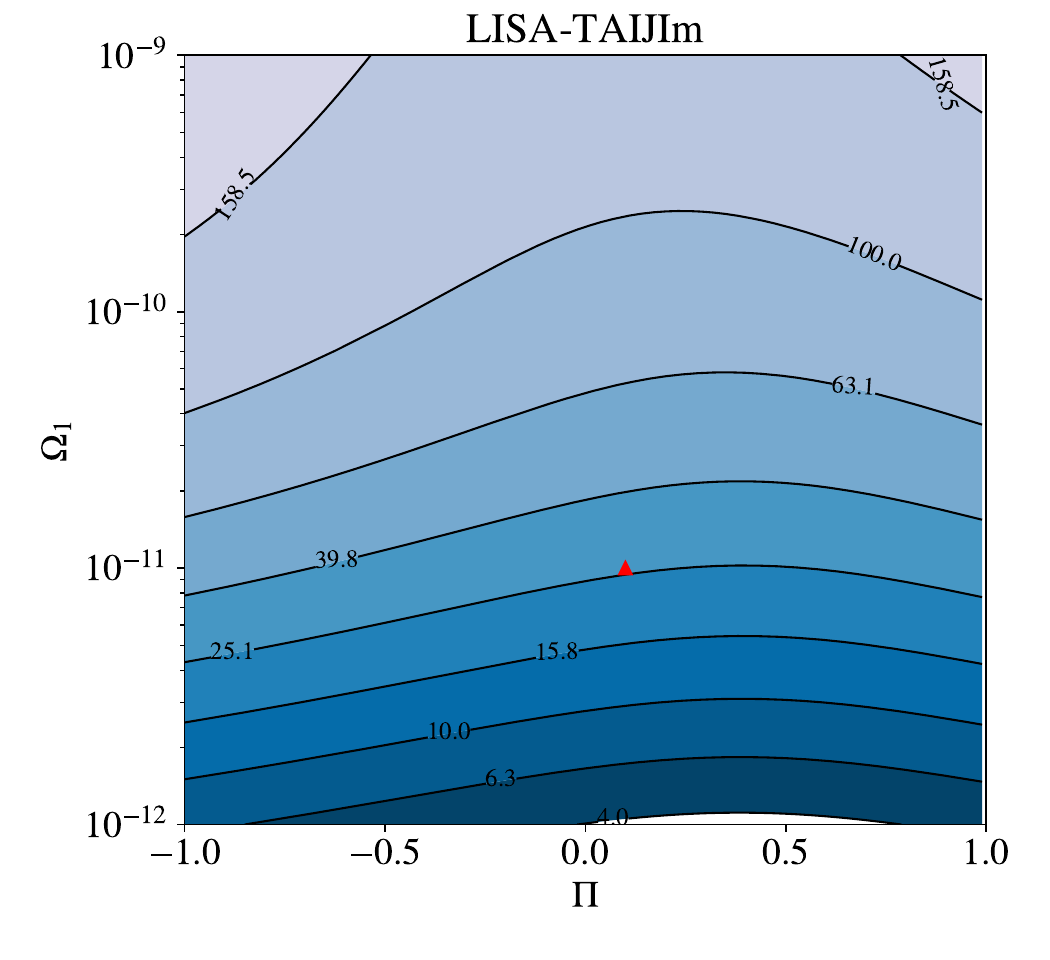} 
 \includegraphics[width=0.45\textwidth]{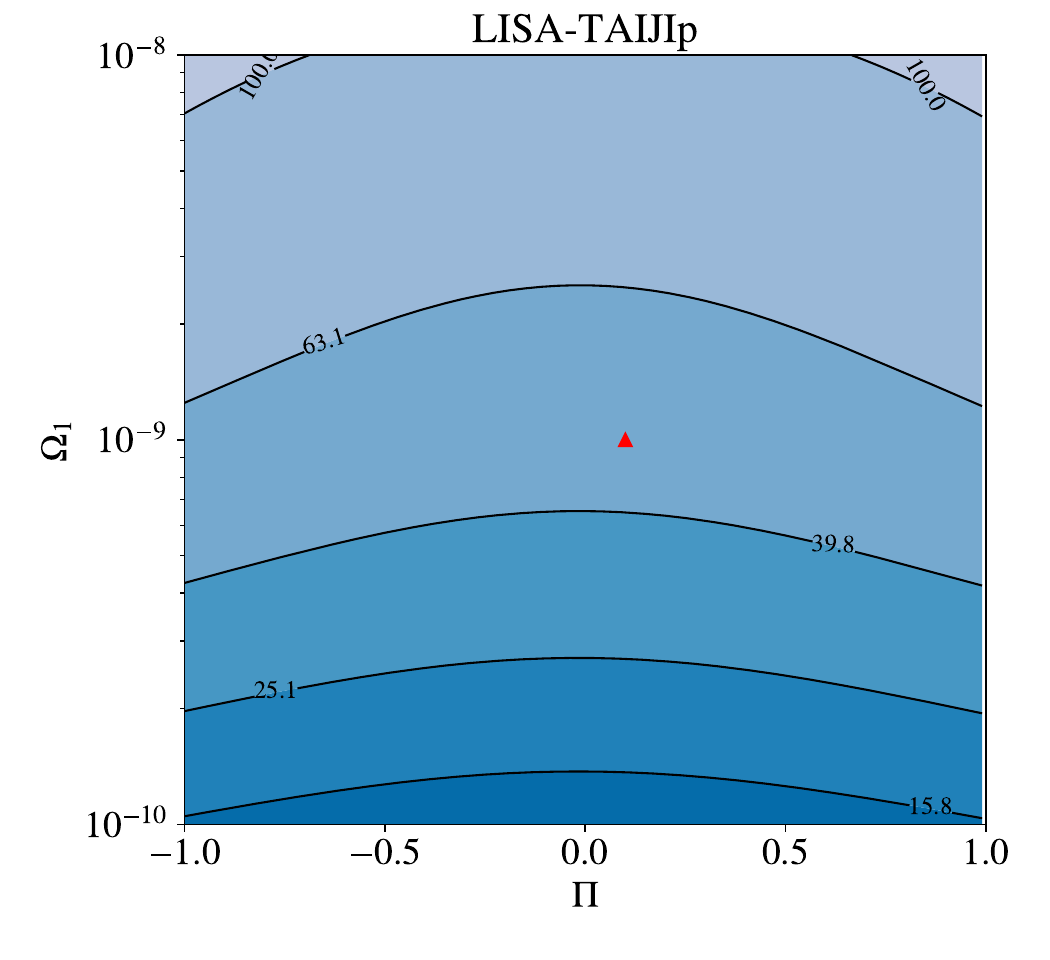}
 \includegraphics[width=0.45\textwidth]{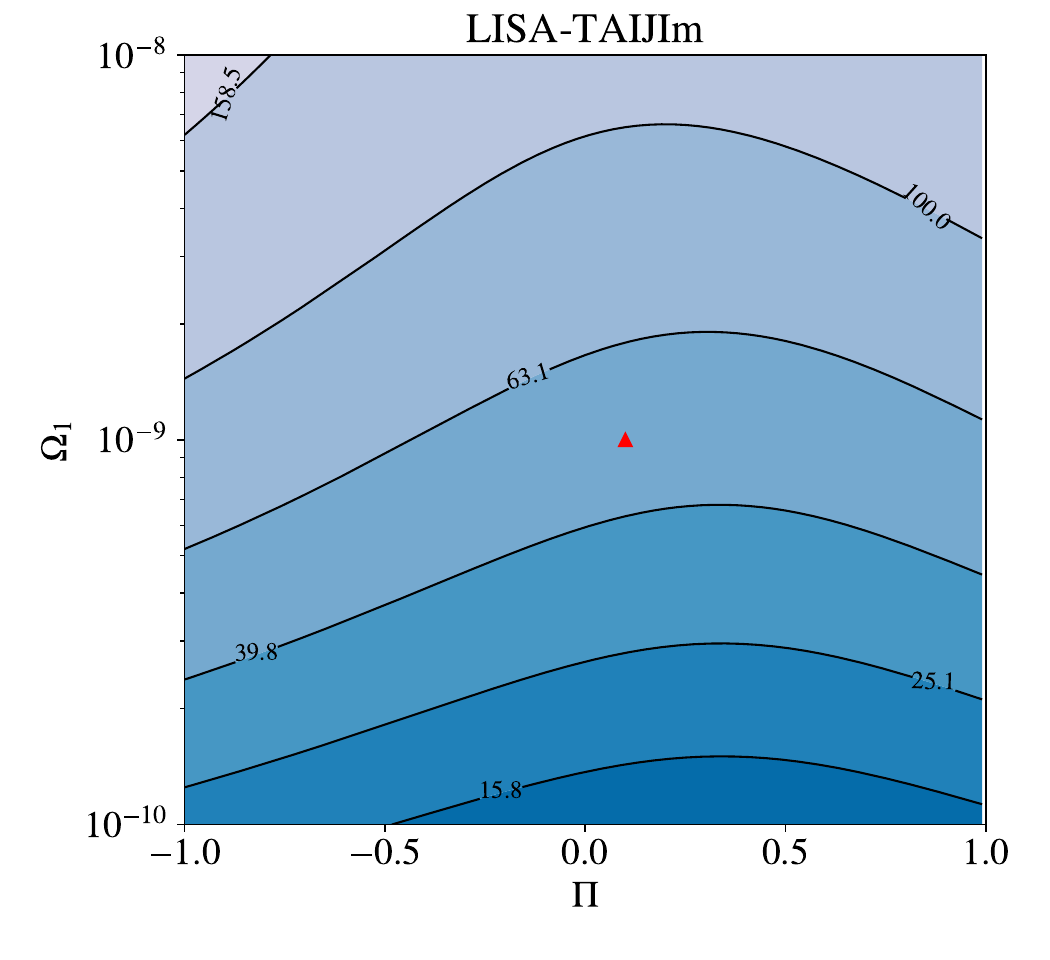}
 \caption{The SNRs for the SGWB with power-law (top), single-peak (middle), and broken power-law (bottom) spectrum models using the LISA-TAIJI networks. The left column displays the SNRs from LISA-TAIJIp, while the right column corresponds to LISA-TAIJIm. The red triangles indicate the fiducial values specified in Eqs.~\eqref{eq:PL_signal}--\eqref{eq:BPL_signal}, with $\Pi = 0.1$.}
 \label{fig:SNRs_LISA_TAIJI}
\end{figure*}

\subsection{Comparisons of SNR from alternative LISA-TAIJI networks}

\begin{figure*}[ht]
 \includegraphics[width=0.45\textwidth]{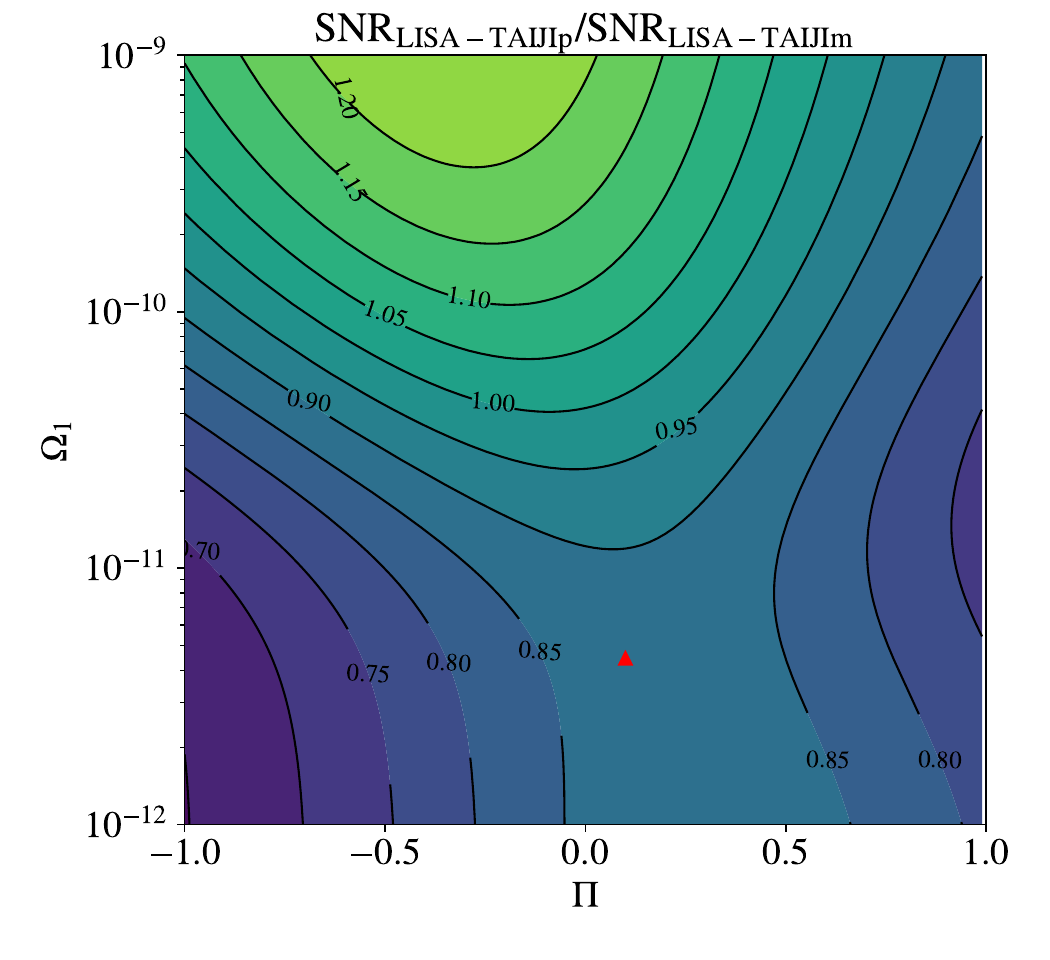}
 \includegraphics[width=0.45\textwidth]{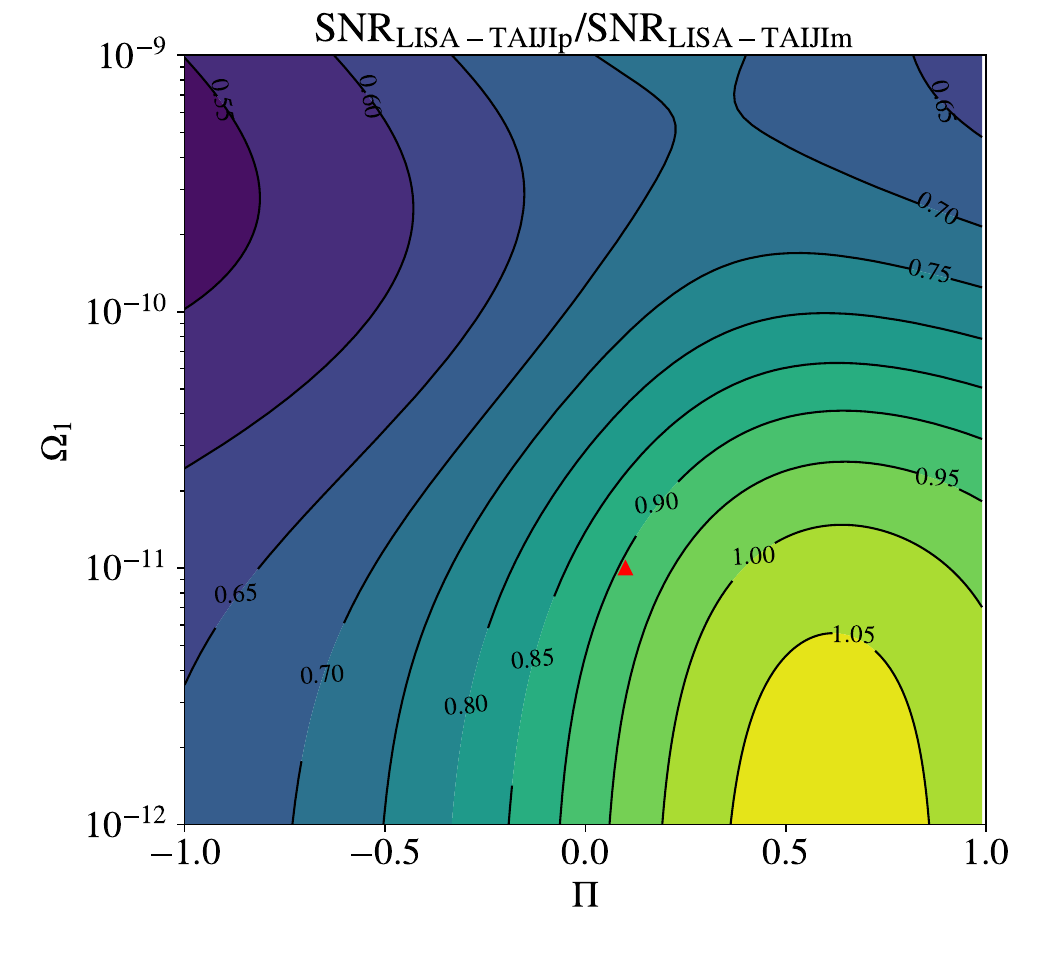} 
 \includegraphics[width=0.45\textwidth]{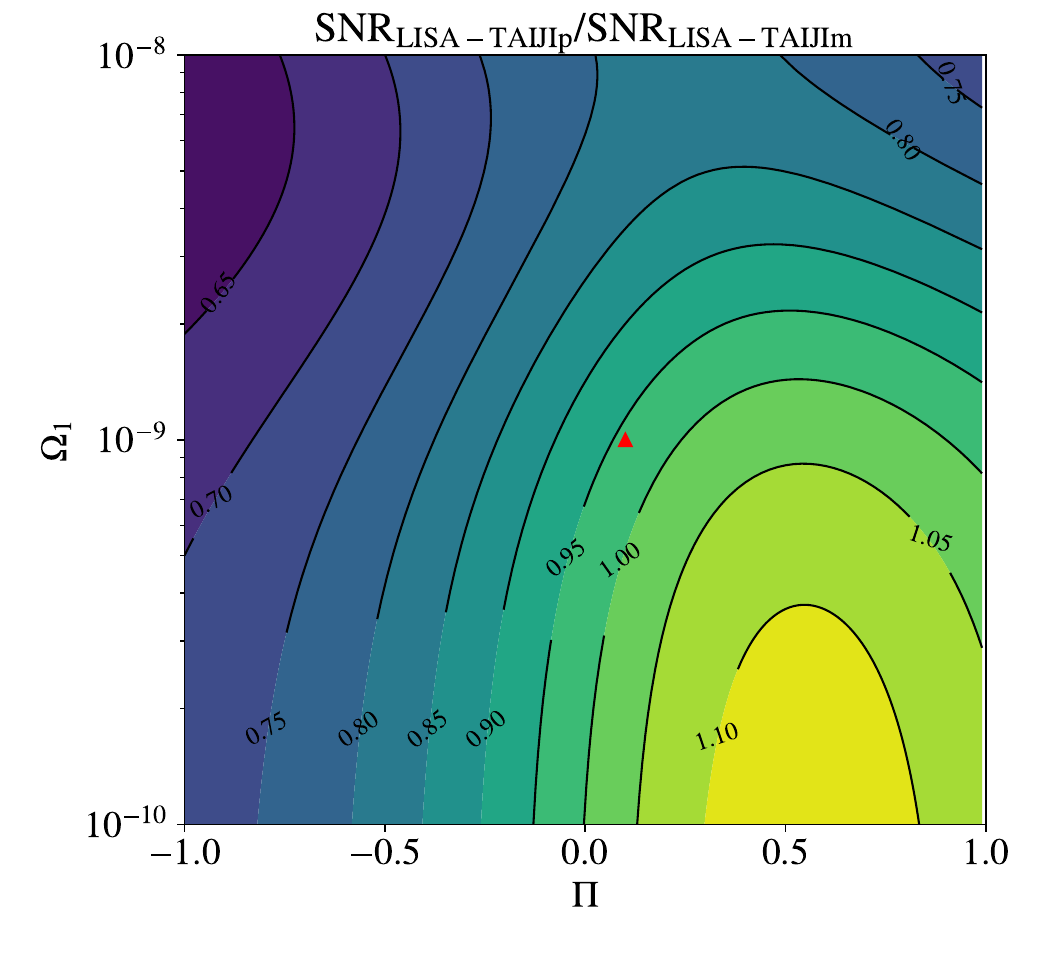}
\caption{The SNR ratios of the two LISA-TAIJI networks for the SGWB with power-law (upper left), single peak (upper right), and broken power-law (lower) energy density spectrum spectrum. In regions of parameter space where the ratio contours exceed 1, the LISA-TAIJIp configuration achieves a higher SNR than LISA-TAIJIm. Conversely, LISA-TAIJIm is more sensitive in regions where the ratio is less than 1. The red triangles indicate the fiducial values specified in Eqs.~\eqref{eq:PL_signal}-\eqref{eq:BPL_signal}, with $\Pi=0.1$.}
 \label{fig:SNRs_ratios}
\end{figure*}

To compare the total SNRs of the two LISA-TAIJI networks across the parameter space, we treat the amplitude $\Omega_1$ and the polarization parameter $\Pi$ as adjustable, while keeping other parameters fixed at their fiducial values for each model. The SNR is computed by varying $\Omega_1$ and $\Pi$ within predefined ranges. The resulting SNRs for the two networks are shown in Fig.~\ref{fig:SNRs_LISA_TAIJI}, while Fig.~\ref{fig:SNRs_ratios} presents the SNR ratios, highlighting the relative sensitivity of the two configurations.

For the power-law and single-peak models, we set $\Omega_1 \in [10^{-12}, 10^{-9}]$ and $\Pi \in [-1, 1]$. LISA-TAIJIm shows lower sensitivity than LISA-TAIJIp at higher amplitudes and lower values of $\Pi$. However, due to its enhanced sensitivity to the $V$ component, LISA-TAIJIm outperforms LISA-TAIJIp for a more highly polarized SGWB. For the single-peak spectrum, the LISA-TAIJIp configuration achieves a higher SNR at lower amplitudes and $\Pi \sim 0.6$. In other regions of the parameter space, LISA-TAIJIm exhibits better sensitivity to the SGWB.

For the broken power-law spectrum, we adpot $\Omega_1 \in [10^{-10}, 10^{-8}]$ and $\Pi \in [-1, 1]$. The trends are similar to those observed for the single-peak spectrum: LISA-TAIJIp achieves a higher SNR at lower amplitudes and near $\Pi \sim 0.6$, while LISA-TAIJIm provides better sensitivity across the rest of the parameter space.

\subsection{Determining parameters of SGWB}

\begin{figure*}[ht]
 \includegraphics[width=0.45\textwidth]{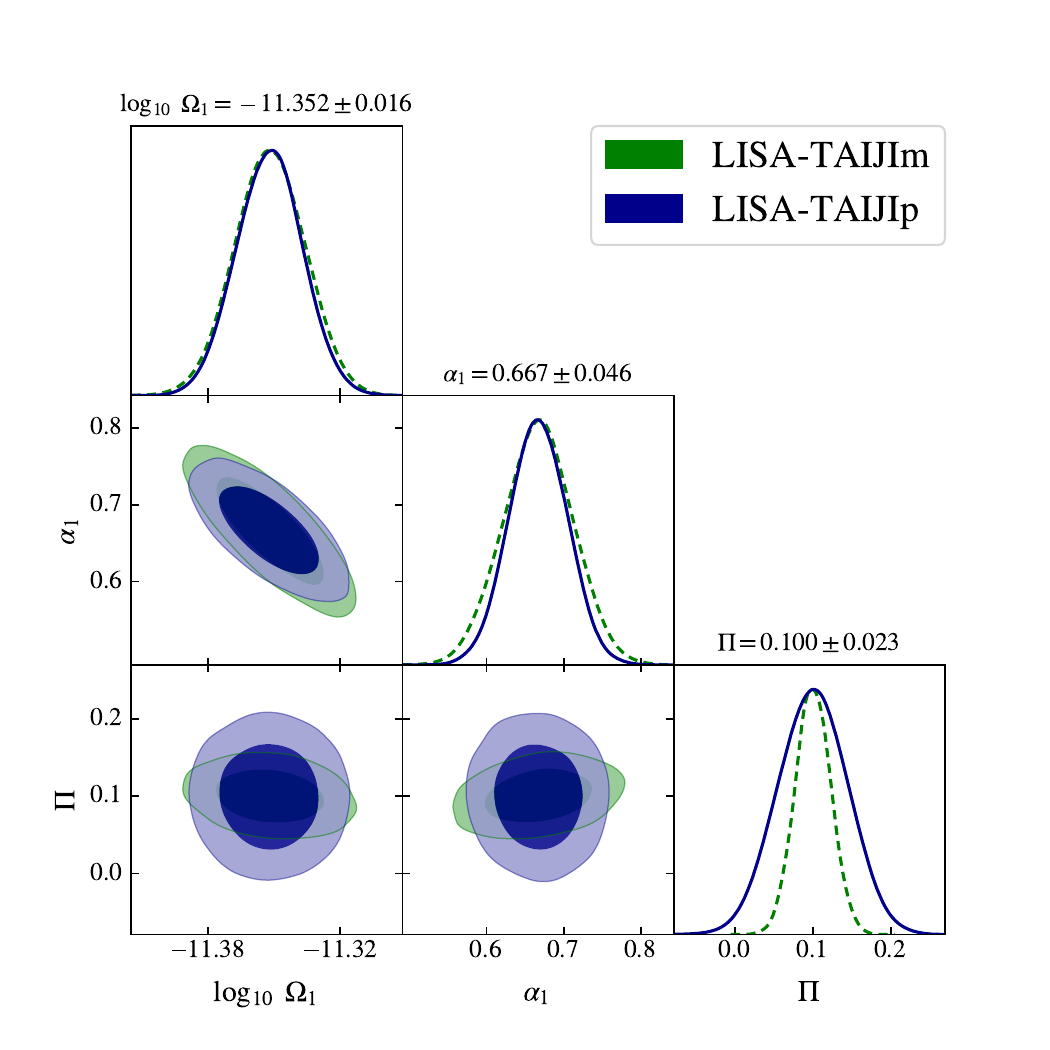}
 \includegraphics[width=0.45\textwidth]{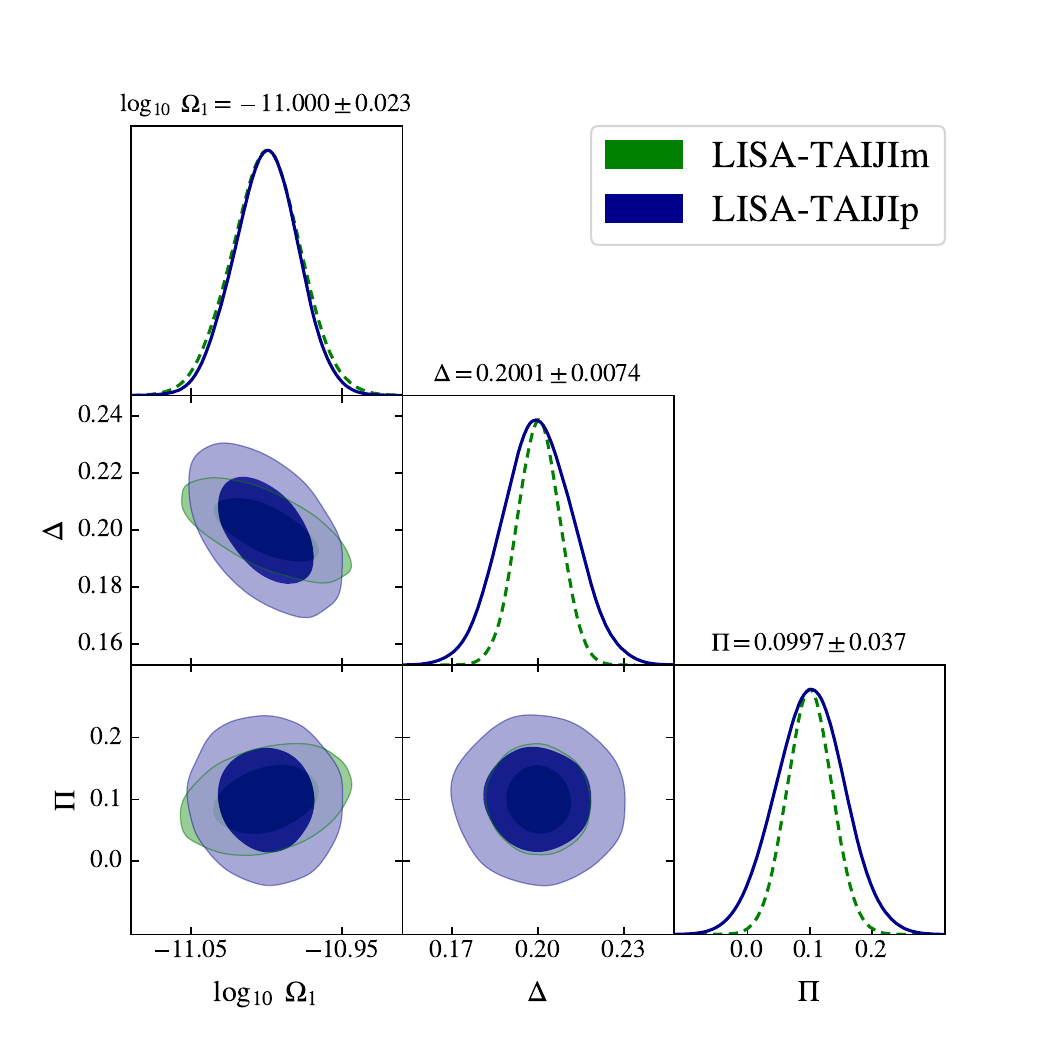} 
 \includegraphics[width=0.60\textwidth]{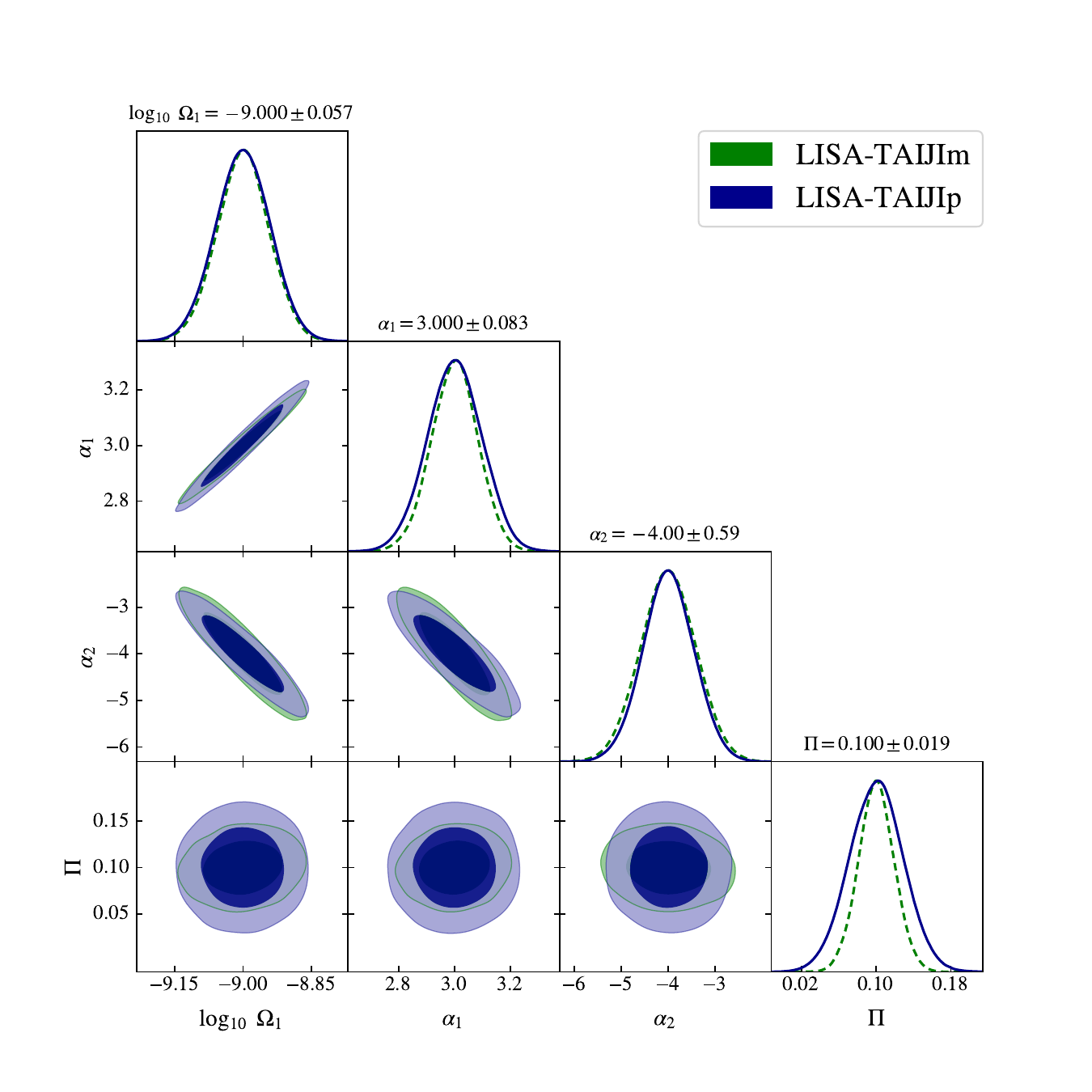} 
 \caption{Corner plots for the parameter uncertainties of SGWB estimated from Fisher information matrix. The upper left plot corresponds to a power-law energy density spectrum with parameters $\Omega_1 = 4.446 \times 10^{-12}$, $\alpha_1 = 2/3$ in Eq.~\eqref{eq:PL_signal} and $\Pi=0.1$. The upper right plot represents a single-peak SGWB with parameters $\Delta = 0.2 $, $\Omega_1 = 1 \times 10^{-11}$ in Eq.~\eqref{eq:SP_signal} and $\Pi=0.1$. The lower plot is for a broken power-law spectrum with parameters $ \alpha_1 = 3, \alpha_2 = -4$, $\Delta =2 $, $\Omega_1 = 1 \times 10^{-9}$ in Eq.~\eqref{eq:BPL_signal} and $\Pi=0.1$. On top of each column, the $1\sigma$ uncertainty of the corresponding parameter is displayed for the LISA-TAIJIm network. }
 \label{fig:corner_plots}
\end{figure*}

In this subsection, we use the Fisher information matrix (FIM) to forecast and compare the constraints on the parameters of different SGWB spectra for two LISA-TAIJI networks~\citep[and references therein]{Kuroyanagi:2018csn, Martinovic:2020hru, Boileau2021, Smith2019a}. The FIM is computed as follows:
\begin{equation} \label{eq:FIM}
\begin{aligned}
 F_{ab} \simeq & \sum_{\kappa} 2 T_\mathrm{obs} \int_{0}^{\infty} \mathrm{d} f \, \frac{ \frac{\partial \langle{C}_{\kappa}\rangle}{\partial \theta_a} \frac{\partial \langle{C}_{\kappa}\rangle}{\partial \theta_b} }{ M_{\kappa}(f) },
\end{aligned}
\end{equation}
where $\kappa$ denotes the channel pairs. A caveat for using the FIM is that the SNR should be sufficiently high for reliable parameter estimation, which implies that $M_{\kappa}$ should take the form given in Eq.~\eqref{eq:M_noApprox}.
The standard deviation of the parameter $\theta_i$ is then given by
\begin{equation}
    \sigma_i \overset{{\rho} \gg 1}{\simeq} \sqrt{\left( F^{-1}_{ab} \right)_{ii}} + \mathcal{O}(\rho^{-1}).
\end{equation}

For the FIM estimation, we use the three SGWB models with fiducial parameters, setting the polarization parameter $\Pi = 0.1$ to represent a slight circular polarization in all three cases. With these parameters, the SNRs from the LISA-TAIJI joint observation are sufficiently high, as indicated by the red triangles in Fig.~\ref{fig:SNRs_LISA_TAIJI}. The SNRs for the power-law, single-peak, and broken power-law models are approximately 40, 25, and 50, respectively.

It is worth noting that these parameter choices may be less favorable for LISA-TAIJIp, where the SNR is lower compared to LISA-TAIJIm, as indicated by the red triangles in Fig.~\ref{fig:SNRs_ratios}. However, reversing or balancing the negative factor for LISA-TAIJIp is challenging, as the favorable parameter space for LISA-TAIJIp is relatively narrow. For instance, in the power-law case, the amplitude $\Omega_1$ needs to be an order of magnitude higher than the fiducial value for LISA-TAIJIp to achieve a comparable or higher SNR than LISA-TAIJIm.

With these fiducial setups, the corner plots for the SGWB parameters of the three spectrum models are shown in Fig.~\ref{fig:corner_plots}. For the power-law model, the uncertainties of the three parameters ($\Omega_1$, $\alpha_1$, $\Pi$) are estimated, as shown in the upper-left plot. And the $1\sigma$ uncertainties for each parameter from LISA-TAIJIm observation are shown at the top of each column. The measurement precisions for $\Omega_1$ and $\alpha_1$ are comparable for two networks, but the polarization parameter $\Pi$ is more precisely determined by the LISA-TAIJIm configuration. Similar trends are observed for the single-peak and broken power-law models, as shown in the upper-right and lower plots of Fig.~\ref{fig:corner_plots}, respectively. While the spectral shape parameters are constrained with comparable precision, the polarization parameter $\Pi$ exhibits smaller uncertainties with LISA-TAIJIm than with LISA-TAIJIp. This improvement can be attributed to the enhanced sensitivity of the LISA-TAIJIm configuration to the $V$ component, as shown in Fig.~\ref{fig:ORF}.

\section{Conclusion and Discussion} \label{sec:conclusion}

The parity violation of the SGWB manifests as an asymmetry between the left-handed and right-handed polarization components, resulting in a circularly polarized SGWB, quantified by the Stokes parameter $V$. A significant degree of circular polarization can arise from various parity-violating mechanisms in the early universe. For instance, during inflation, axion-like fields can amplify tensor perturbations in one polarization state, leading to a fully polarized SGWB~\cite{Machado:2019xuc,Xu:2024kwy}. The detection of such a parity-violating SGWB presents a unique opportunity to explore new physics in the early universe.

To constrain the underlying mechanisms, it is essential to isolate the circular polarization component, $V$. In space-based detectors, the circular polarization of an isotropic SGWB typically cancels due to the mirror symmetry of their planar structures, making them insensitive to the chirality of gravitational waves. However, the LISA-TAIJI networks are not subject to such cancellations, allowing them to potentially detect the parity-violating $V$ component and providing a valuable opportunity to test the associated physical theories.

In this work, we investigate the detectability of circular polarization in an isotropic SGWB using two distinct LISA-TAIJI network configurations. The first configuration, LISA-TAIJIp, consists of two detectors with constellations both inclined at $+60^\circ$ relative to the ecliptic plane. The second configuration, LISA-TAIJIm, pairs LISA with TAIJIm, which is inclined at $-60^\circ$. While the LISA-TAIJIp configuration exhibits slightly better sensitivity to the intensity of the SGWB, the LISA-TAIJIm configuration significantly improves sensitivity to the circular polarization $V$ component, particularly in the low-frequency band, by approximately one order of magnitude.

To quantify the capabilities of the two networks, we evaluate and compare the SNRs for SGWB with power-law, single-peak, and broken power-law energy density spectrum. The results show that LISA-TAIJIm can achieve higher SNRs than LISA-TAIJIp across a broad parameter space, thanks to its improved sensitivity to the circular polarization component. Additionally, we employ the Fisher information matrix (FIM) to forecast the precision in parameter estimation of the energy density spectrum. The FIM results confirm that LISA-TAIJIm provides better constraints on the circular polarization parameter $\Pi$, offering a great opportunity to test parity-violating theories in the milli-Hz frequency band.

In conclusion, the parity-violating $V$ component of an isotropic SGWB can be detected by space-borne GW detector networks. We compared the detectability of such $V$-component signals for different configurations and found that the sensitivity to the $V$ component in low frequency range can be improved by approximately one order of magnitude using the LISA-TAIJIm network, compared to LISA-TAIJIp. And more detailed analysis shows that LISA-TAIJIm offers better sensitivity across a broad parameter space and provides tighter constraints on the circular polarization parameter $\Pi$. This presents a promising opportunity to test various parity-violating theories in the mHz band.

\begin{acknowledgments}
This work is supported by the National Key Research and Development Program of China under Grants (No.2021YFC2201903, 2021YFC2201901), the National Natural Science Foundation of China under Grants (No. 12375059, No. 12147132, No. 12405074, No. 12347103), and the Fundamental Research Funds for the Central Universities (No. E2ET0209X2). This work made use of the High-Performance Computing Resource in the Core Facility for Advanced Research Computing at Shanghai Astronomical Observatory. The calculations in this work are performed by using the python packages $\mathsf{numpy}$ \cite{harris2020array} and $\mathsf{scipy}$ \cite{2020SciPy-NMeth}, and the plots are made by utilizing $\mathsf{matplotlib}$ \cite{Hunter:2007ouj} and $\mathsf{GetDist}$ \cite{Lewis:2019xzd}. We thank helpful discussions with Jing Liu, Yong Tang and Yang Jiang. 
\end{acknowledgments}

\bibliography{stochastic}

\end{document}